\documentclass[a4paper,fleqn,usenatbib]{mnras}

\usepackage{amsmath}	% Advanced maths commands
\usepackage{amssymb}	% Extra maths symbols
\usepackage{txfonts}

\usepackage[T1]{fontenc}
\usepackage{ae,aecompl}

\usepackage{graphicx}	% Including figure files
\usepackage{placeins}

\newcommand{\AV}{\hbox{$A_{\rm V}$}}
\newcommand{\BJDc}{\hbox{BJD$_{\rm c}$}}
\newcommand{\Bl}{\hbox{$B_{\rm \ell}$}}
\newcommand{\BmV}{\hbox{${\rm B-V}$}}
\newcommand{\chisqr}{\hbox{$\chi^2_{\rm r}$}}
\newcommand{\dom}{\hbox{$d\Omega$}}
\newcommand{\kms}{\hbox{km\,s$^{-1}$}}
\newcommand{\logg}{\hbox{$\log g$}}
\newcommand{\logL}{\hbox{$\log($\lstar/\lsun$)$}}
\newcommand{\lstar}{\hbox{$L_{\star}$}}
\newcommand{\lsun}{\hbox{${\rm L}_{\odot}$}}
\newcommand{\mjup}{\hbox{${\rm M}_{\rm Jup}$}}
\newcommand{\mps}{\hbox{m\,s$^{-1}$}}
\newcommand{\msini}{\hbox{$M\sin i$}}
\newcommand{\mstar}{\hbox{$M_{\star}$}}
\newcommand{\msun}{\hbox{${\rm M}_{\odot}$}}
\newcommand{\omeq}{\hbox{$\Omega_{\rm eq}$}}
\newcommand{\popr}{\hbox{$P_{\rm orb}/P_{\rm rot}$}}
\newcommand{\Porb}{\hbox{$P_{\rm orb}$}}
\newcommand{\Prot}{\hbox{$P_{\rm rot}$}}
\newcommand{\rpd}{\hbox{rad\,d$^{-1}$}}
\newcommand{\rstar}{\hbox{$R_{\star}$}}
\newcommand{\rsun}{\hbox{${\rm R}_{\odot}$}}
\newcommand{\rvfil}{\hbox{RV$_{\rm filt}$}}
\newcommand{\rvraw}{\hbox{RV$_{\rm raw}$}}
\newcommand{\sn}{\hbox{S/N}}
\newcommand{\teff}{\hbox{$T_{\rm eff}$}}
\newcommand{\V}{\hbox{${\rm V}$}}
\newcommand{\VmRj}{\hbox{${\rm V-R_{\rm J}}$}}
\newcommand{\vrad}{\hbox{$v_{\rm rad}$}}
\newcommand{\vsini}{\hbox{$v \sin i$}}
\newcommand{\dchis}{\hbox{$\Delta\chi^2$}}
\newcommand{\logLr}{\hbox{$\log\mathcal{L}_r$}}
\newcommand{\caii}{\hbox{Ca$\;${\sc ii}}}
\newcommand{\fei}{\hbox{Fe$\;${\sc i}}}
\newcommand{\hei}{\hbox{He$\;${\sc i}}}
\newcommand{\hal}{\hbox{H${\alpha}$}}

\title[A hot Jupiter around the active wTTS TAP~26]{A hot Jupiter around the very active weak-line T~Tauri star TAP~26}

\author[L.~Yu et al.]{L.~Yu$^{1,2}$\thanks{E-mail: louise.yu@irap.omp.eu}, J.-F.~Donati$^{1,2}$, E.~M.~H\'ebrard$^3$, C.~Moutou$^4$, L.~Malo$^5$, K.~Grankin$^6$,
	\newauthor G.~Hussain$^{7,1}$, A.~Collier Cameron$^8$, A.~A.~Vidotto$^9$, C.~Baruteau$^{1,2}$, S.H.P.~Alencar$^{10}$,
	\newauthor J.~Bouvier$^{11,12}$, P.~Petit$^{1,2}$, M.~Takami$^{13}$, G.~Herczeg$^{14}$, S.~G.~Gregory$^8$, M.~Jardine$^8$,
	\newauthor J.~Morin$^{15}$, F.~M\'enard$^{11,12}$ and the MaTYSSE collaboration\\\\
	All affiliations are listed at the end of the paper
}

\date{Accepted XXX. Received YYY; in original form ZZZ}

\pubyear{2016}

\begin{document}
	\label{firstpage}
	\pagerange{\pageref{firstpage}--\pageref{lastpage}}
	\maketitle
	
	\begin{abstract}
		We report the results of an extended spectropolarimetric and photometric monitoring of the weak-line T Tauri star TAP~26, carried out within the MaTYSSE programme with the ESPaDOnS spectropolarimeter at the 3.6~m Canada-France-Hawaii Telescope. Applying Zeeman-Doppler Imaging to our observations, concentrating in 2015 November and 2016 January and spanning 72~d in total, 16~d in 2015 November and 13~d in 2016 January, we reconstruct surface brightness and magnetic field maps for both epochs and demonstrate that both distributions exhibit temporal evolution not explained by differential rotation alone. We report the detection of a hot Jupiter (hJ) around TAP~26 using three different methods, two using Zeeman-Doppler Imaging (ZDI) and one Gaussian-Process Regression (GPR), with a false-alarm probability smaller than $6\ 10^{-4}$. However, as a result of the aliasing related to the observing window, the orbital period cannot be uniquely determined; the orbital period with highest likelihood is 10.79$\pm$0.14~d followed by 8.99$\pm$0.09~d. Assuming the most likely period, and that the planet orbits in the stellar equatorial plane, we obtain that the planet has a minimum mass \msini\ of 1.66$\pm$0.31~\mjup\ and orbits at 0.0968$\pm$0.0032~au from its host star. This new detection suggests that disc type II migration is efficient at generating newborn hJs, and that hJs may be more frequent around young T~Tauri stars than around mature stars (or that the MaTYSSE sample is biased towards hJ-hosting stars).
	\end{abstract}
	
	\begin{keywords}
		magnetic fields --  
		planets and satellites: formation -- 
		stars: imaging -- 
		stars: rotation -- 
		stars: individual:  TAP~26 --
		techniques: polarimetric
	\end{keywords}
	
	%%%%%%%%%%%%%%%%%%%%%%%%%%%%%%%%%%%%%%%%%%%%%%%%%%
	
	%%%%%%%%%%%%%%%%% BODY OF PAPER %%%%%%%%%%%%%%%%%%
	
	\section{Introduction}
	\label{sec:int}
	
	Studying young forming stars stands as our best chance to progress in our understanding of the formation and early evolution of planetary systems. For instance, detecting hot Jupiters (hJs) around young stars (1-10~Myrs) and determining their orbital properties can enable us to clarify how they form and migrate, and to better characterise the physical processes (e.g. planet-disc interaction, planet-planet scattering, \citealt{Baruteau14}, in-situ formation, \citealt{Batygin16}) responsible for generating such planets.
	\begin{table*}
		\centering
		\caption[]{Journal of ESPaDOnS observations of TAP~26 collected in 2015~November (first 16 lines) and 2016~January (last 13 lines). Each observation consists of a sequence of 4 subexposures, each lasting 695~s. Columns~1 to 4 respectively list (i) the UT date of the observation, (ii) the corresponding UT time (at mid-exposure), (iii) the Barycentric Julian Date (BJD) in excess of 2,457,300, and (iv) the peak signal to noise ratio (per 2.6~\kms\ velocity bin) of each observation. Column~5 lists the root-mean-square (rms) noise level (relative to the unpolarised continuum level $I_{\rm c}$ and per 1.8~\kms\ velocity bin) in the circular polarisation profiles produced by Least-Squares Deconvolution (LSD) and column~6 lists the signal to noise ratio in the unpolarised profiles produced by LSD, measured from the noise level in intervals of continuum of the LSD profiles. Column~7 indicates the rotational cycle associated with each exposure (using the ephemeris given by Eq.~\ref{eq:eph}). Column~8 lists the raw RVs computed from the unpolarised spectra, column~9 the filtered RVs (see Sec.~\ref{sec:fil}) and column~10 the 1$\sigma$ error bar on both \rvraw\ and \rvfil. Column 11-13 list values for activity proxies mentioned in App.~\ref{sec:act}: the line-of-sight-projected magnetic field averaged over the visible stellar hemisphere (also called longitudinal field) and the equivalent width of the \hal\ emission (counted from above the continuum level, expressed in \kms, and with a typical 1$\sigma$ error bar of 3.0~\kms) .}
		\begin{tabular}{ccccccccccccc}
			\hline
			Date & UT & BJD & \sn & $\sigma_{\rm LSD}$ & \sn$_{I}$ & Cycle & \rvraw & \rvfil & $\sigma_{\rm RV}$ & \Bl & $\sigma_{B_{\rm \ell}}$ & EW$_{\rm H \it \alpha}$ \\
			& (h:m:s) & (2,457,300+) & & ($10^{-4}$) & & & (\kms) & (\kms) & (\kms) & (G) & (G) & (\kms) \\
			\hline
			18 Nov & 09:36:28 & 44.90594 & 140 & 3.3 & 1867 & 0.148 & 1.049 & 0.141 & 0.075 & 99 & 45 & 39.3 \\
			22 Nov & 12:11:18 & 49.01352 & 140 & 3.3 & 1835 & 5.905 & -1.115 & 0.026 & 0.076 & -72 & 47 & 37.6 \\
			23 Nov & 11:20:34 & 49.97830 & 140 & 3.1 & 1862 & 7.258 & 0.677 & -0.120 & 0.075 & -20 & 46 & 36.2 \\
			24 Nov & 11:20:25 & 50.97819 & 140 & 3.0 & 1890 & 8.659 & 0.915 & -0.020 & 0.074 & -143 & 45 & 43.1 \\
			25 Nov & 07:41:04 & 51.82588 & 140 & 3.3 & 1804 & 9.847 & -0.017 & -0.149 & 0.078 & -182 & 47 & 44.0 \\
			25 Nov & 13:49:53 & 52.08201 & 140 & 3.2 & 1861 & 10.206 & 1.204 & -0.077 & 0.075 & -28 & 46 & 29.3 \\
			26 Nov & 10:09:09 & 52.92871 & 150 & 3.0 & 1922 & 11.393 & -0.791 & -0.176 & 0.073 & 71 & 44 & 26.9 \\
			27 Nov & 11:36:33 & 53.98941 & 120 & 3.9 & 1866 & 12.879 & -0.590 & -0.087 & 0.075 & -44 & 46 & 52.7 \\
			28 Nov & 11:25:28 & 54.98171 & 110 & 4.0 & 1849 & 14.270 & 0.491 & -0.019 & 0.076 & -59 & 46 & 37.4 \\
			29 Nov & 08:19:32 & 55.85260 & 140 & 3.1 & 1894 & 15.491 & 0.224 & -0.016 & 0.074 & 26 & 45 & 38.7 \\
			29 Nov & 11:15:55 & 55.97508 & 140 & 3.3 & 1870 & 15.662 & 1.007 & 0.052 & 0.075 & -129 & 46 & 42.1 \\
			30 Nov & 07:30:58 & 56.81887 & 150 & 3.2 & 1863 & 16.845 & 0.508 & 0.184 & 0.075 & -199 & 46 & 44.7 \\
			01 Dec & 08:19:49 & 57.85279 & 140 & 3.2 & 1879 & 18.294 & 0.273 & 0.187 & 0.075 & -107 & 45 & 47.2 \\
			01 Dec & 11:18:25 & 57.97681 & 130 & 3.4 & 1909 & 18.468 & 0.158 & 0.084 & 0.074 & 40 & 45 & 44.1 \\
			02 Dec & 07:48:41 & 58.83116 & 150 & 3.1 & 1887 & 19.665 & 1.068 & 0.097 & 0.074 & -164 & 45 & 45.9 \\
			03 Dec & 09:55:37 & 59.91929 & 150 & 3.0 & 1899 & 21.190 & 1.147 & 0.082 & 0.074 & 51 & 45 & 30.4 \\
			\hline
			17 Jan & 09:19:04 & 104.89186 & 130 & 3.5 & 1759 & 84.221 & 0.200 & -0.070 & 0.080 & -45 & 49 & 34.0 \\
			18 Jan & 05:01:52 & 105.71318 & 140 & 3.2 & 1816 & 85.372 & -0.500 & -0.144 & 0.077 & -15 & 47 & 24.5 \\
			19 Jan & 05:02:31 & 106.71356 & 140 & 3.4 & 1772 & 87.774 & 0.594 & -0.140 & 0.079 & -36 & 48 & 57.9 \\
			20 Jan & 07:55:33 & 107.83363 & 100 & 4.8 & 1708 & 88.344 & -0.478 & -0.078 & 0.082 & -48 & 50 & 26.6 \\
			21 Jan & 05:04:22 & 108.71467 & 140 & 3.4 & 1792 & 89.579 & 0.613 & -0.067 & 0.078 & 71 & 48 & 37.6 \\
			22 Jan & 05:04:03 & 109.71438 & 120 & 4.1 & 1738 & 90.980 & -0.937 & 0.068 & 0.081 & -201 & 49 & 44.0 \\
			23 Jan & 06:06:31 & 110.75767 & 140 & 3.3 & 1802 & 92.442 & 0.376 & 0.190 & 0.078 & 1 & 47 & 38.9 \\
			24 Jan & 05:05:28 & 111.71519 & 140 & 3.2 & 1780 & 93.784 & 0.944 & 0.102 & 0.079 & -127 & 48 & 46.4 \\
			25 Jan & 06:30:41 & 112.77428 & 140 & 3.3 & 1805 & 95.269 & -0.014 & 0.169 & 0.078 & 27 & 47 & 37.0 \\
			26 Jan & 06:03:54 & 113.75560 & 140 & 3.5 & 1767 & 96.644 & 0.778 & 0.100 & 0.079 & -51 & 48 & 44.9 \\
			27 Jan & 06:58:50 & 114.79365 & 140 & 3.4 & 1774 & 98.099 & -1.185 & -0.011 & 0.079 & -2 & 48 & 39.7 \\
			28 Jan & 06:59:12 & 115.79383 & 140 & 3.4 & 1737 & 99.501 & 0.548 & -0.019 & 0.081 & 70 & 49 & 41.8 \\
			29 Jan & 06:05:30 & 116.75644 & 130 & 3.5 & 1758 & 100.850 & 0.958 & 0.062 & 0.080 & -71 & 49 & 41.7 \\
			\hline
		\end{tabular}
		\label{tab:log}
	\end{table*}
	
	However, young stars are enormously active, rendering planet signatures in their spectra and / or light-curves extremely difficult to detect in practice. Until very recently, most planets found so far around stars younger than 20~Myr were distant planets detected with imaging techniques (e.g. $\beta$~Pic~b, \citealt{Lagrange10}, and LkCa~15, \citealt{Sallum15}). Early claims of hJs orbiting around T Tauri stars \citep[e.g. TW~Hya,][]{Setiawan08} finally proved to be activity signatures mistakenly interpreted as radial velocity (RV) signals from close-in giant planets \citep{Huelamo08}.  
	
	Following the recent discovery of newborn close-in giant planets \citep{Mann16,David16,Donati16a} or planet candidates \citep{Eyken12,JohnsKrull16} around forming stars, time is ripe for a systematic exploration of hJs around T Tauri stars, and in particular the so called weak-line T Tauri stars (wTTSs), whose accretion disc has just dissipated. This is one of the main goals of the MaTYSSE (Magnetic Topologies of Young Stars and the Survival of close-in massive Exoplanets) large-programme allocated on the 3.6m Canada-France-Hawaii Telescope (CFHT), thanks to which the youngest hJ discovered so far was detected \citep{Donati16a,Donati17} and within which this study places.
	
	In this paper, we present results for another wTTS, the young pre-main sequence (PMS) solar-mass star, TAP~26, \citep{Feigelson87,Grankin08,Grankin13}, located in the Taurus star-forming region. TAP~26 was observed in late 2015 and early 2016 with both the ESPaDOnS spectropolarimeter and the 1.25~m telescope at the Crimean Astrophysical Observatory (CrAO). After documenting our observations (Sec.~\ref{sec:obs}), we derive the stellar parameters of TAP~26 (Sec.~\ref{sec:evo}), before reconstructing the surface magnetic and brightness maps by applying Zeeman-Doppler Imaging (ZDI) to our data (Sec.~\ref{sec:mod}). We finally detail in Sec.~\ref{sec:mpl} our detection of a planet radial velocity (RV) signal in its spectrum, using three different methods. The first two methods are based on ZDI following previous studies \citep{Donati15,Donati16a,Petit15}, and the third one exploits Gaussian-Process Regression \citep[GPR,][see Sec.~\ref{sec:mpl}]{Haywood14,Rajpaul15}.
	
	\section{Observations}\label{sec:obs}
	\begin{table}
		\centering
		\caption[]{Journal of contemporaneous CrAO multicolour photometric observations of TAP~26 collected in late 2015 and early 2016, respectively listing the UT date and Heliocentric Julian Date (HJD) of the observation, the measured \V\ magnitude (1$\sigma$ error bar of 0.016~mag) and \VmRj\ Johnson photometric colours, and the corresponding rotational phase (using again the ephemeris given by Eq.~\ref{eq:eph}). The table is divided into three periods spanning 1.5-2.5~months each, the second one covering the 2015 Nov set of spectropolarimetric observations and the third one overlapping the 2016 Jan set of spectropolarimetric observations.}
		\begin{tabular}{ccccc}
			\hline
			Date   & HJD      & \V  & \VmRj & Cycle \\
			& (2,457,200+) & (mag) & & (-120+) \\
			\hline
			25 Aug & 60.569 & 12.291 & - & 1.946 \\
			30 Aug & 65.592 & 12.269 & 0.986 & 8.987 \\
			31 Aug & 66.583 & 12.261 & 1.010 & 10.375 \\
			09 Sep & 75.557 & 12.297 & 1.016 & 22.953 \\
			11 Sep & 77.562 & 12.331 & 1.022 & 25.763 \\
			16 Sep & 82.564 & 12.329 & 1.004 & 32.774 \\
			18 Sep & 84.594 & 12.259 & 1.004 & 35.619 \\
			19 Sep & 85.530 & 12.300 & 1.007 & 36.930 \\
			22 Sep & 88.529 & 12.260 & 1.003 & 41.134 \\
			23 Sep & 89.505 & 12.245 & 1.014 & 42.501 \\
			24 Sep & 90.517 & 12.282 & 0.988 & 43.920 \\
			25 Sep & 91.550 & 12.246 & 0.988 & 45.369 \\
			26 Sep & 92.524 & 12.320 & 1.001 & 46.733 \\
			28 Sep & 94.550 & 12.238 & 0.968 & 49.573 \\
			03 Oct & 99.588 & 12.283 & 1.030 & 56.633 \\
			04 Oct & 100.513 & 12.276 & 0.983 & 57.930 \\
			09 Oct & 105.545 & 12.280 & 1.016 & 64.982 \\
			\hline
			15 Oct & 111.600 & 12.232 & 0.967 & 73.469 \\
			16 Oct & 112.605 & 12.292 & 0.976 & 74.877 \\
			17 Oct & 113.595 & 12.269 & 1.000 & 76.265 \\
			19 Oct & 115.597 & 12.261 & 0.984 & 79.070 \\
			20 Oct & 116.584 & 12.233 & 0.963 & 80.454 \\
			25 Oct & 121.564 & 12.263 & 1.014 & 87.434 \\
			27 Oct & 123.507 & 12.247 & 0.994 & 90.157 \\
			30 Oct & 126.442 & 12.280 & 1.024 & 94.270 \\
			03 Nov & 130.564 & 12.220 & 1.012 & 100.048 \\
			13 Nov & 140.585 & 12.229 & 0.989 & 114.092 \\
			16 Dec & 173.373 & 12.245 & 1.003 & 160.046 \\
			17 Dec & 174.306 & 12.238 & 0.979 & 161.354 \\
			\hline
			03 Jan & 191.364 & 12.215 & 0.976 & 185.262 \\
			17 Jan & 205.347 & 12.306 & 0.983 & 204.860 \\
			24 Jan & 212.316 & 12.245 & 1.009 & 214.626 \\
			30 Jan & 218.296 & 12.297 & 1.019 & 223.008 \\
			10 Feb & 229.258 & 12.217 & 0.975 & 238.371 \\
			22 Feb & 241.262 & 12.245 & 0.982 & 255.195 \\
			05 Mar & 253.253 & 12.293 & 0.987 & 272.002 \\
			08 Mar & 256.285 & 12.238 & 0.992 & 276.251 \\
			15 Mar & 263.268 & 12.299 & 1.002 & 286.038 \\
			\hline
		\end{tabular}
		\label{tab:pho}
	\end{table}
	
	TAP~26 was observed in November~2015 and January~2016 using the high-resolution spectropolarimeter ESPaDOnS at the 3.6-m CFHT at Mauna Kea (Hawaii). ESPaDOnS collects stellar spectra spanning the entire optical domain (from 370 to 1,000~nm) at a resolving power of 65,000 (i.e., resolved velocity element of 4.6~\kms) over the full wavelength range \citep{Donati03}. A total of 29 unpolarised (Stokes~$I$) and circularly-polarised (Stokes~$V$) spectra were collected over a timespan of 72~d, 16 spectra over 16 nights in 2015 Nov, and 13 spectra over 13 nights in 2016 Jan. The rate was of one spectrum per night, except at the beginning of the 2015 Nov session where a three-day gap following the first observation was compensated by pairs of observations on Nov 25, Nov 29 and Dec 01. However, given the 0.71~d rotation period of TAP~26, phase coverage is not optimal and the 2015 Nov data set presents gaps of \mbox{0.15-0.25~rotation cycle} (see Table~\ref{tab:log}).
	
	Each polarisation exposure sequence consists of 4 individual subexposures taken in different polarimeter configurations to allow the removal of all spurious polarisation signatures at first order. All raw frames are processed with the nominal reduction package {\sc Libre ESpRIT} as described in the previous papers of the series \citep[e.g.,][]{Donati10b, Donati11, Donati14}, yielding a typical rms RV precision of 20-30~\mps\ \citep{Moutou07,Donati08b}. The peak signal-to-noise ratios (\sn, per 2.6~\kms\ velocity bin) achieved on the collected spectra range between 100 and 150 (median 140), depending mostly on weather/seeing conditions. The full journal of observations is presented in Table~\ref{tab:log}.
	
	Rotational cycles (noted $E$ in the following equation) are computed from Barycentric Julian Dates (BJDs) according to the ephemeris:
	\begin{equation}
	\mbox{BJD} \hbox{\rm ~(d)} = 2,457,344.8 + \Prot E
	\label{eq:eph}
	\end{equation}
	in which the photometrically-determined rotation period \Prot\ \citep[equal to 0.7135~d,][]{Grankin13} is taken from the literature and the initial Julian date (2,457,344.8~d) is chosen arbitrarily.
	
	Least-Squares Deconvolution \citep[LSD,][]{Donati97b} was applied to all spectra. The line list we employed for LSD is computed from an {\sc Atlas9} LTE model atmosphere \citep{Kurucz93} featuring \teff=4,500~K and \logg=4.5, the most appropriate model for TAP~26 (see Sec.~\ref{sec:evo}). Only moderate to strong atomic spectral lines are included in this list \citep[see, e.g.,][for more details]{Donati10b}. Altogether, about 7,800 spectral features (with about 40\% from \fei) are used in this process. The Stokes $I$ and Stokes $V$ LSD profiles can be seen in Sec.~\ref{sec:mod}. Significant distortions are visible in all Stokes~$I$ LSD profiles, indicating the presence of brightness inhomogeneities covering a large fraction of the surface of TAP~26 at the time of our observations. The noise level in Stokes $I$ LSD profiles is measured from continuum intervals (see Table~\ref{tab:log}), and includes not only the noise from photon statistics, but also the (often dominant) noise introduced by LSD.
	
	Among the 29 profiles we used, 11 were contaminated by solar light reflected off the Moon (5 in 2015 Nov, the Moon being at 9.5\degr\ from TAP~26 and at 99\% illumination on 2015 Nov 26, and 6 in 2016 Jan, the Moon being at 12\degr\ from TAP~26 and at 85\% illumination on 2016 Jan 19); we applied a two-step process involving tomographic imaging, described in \citet{Donati16a}, to filter out this contamination from our Stokes~$I$ LSD profiles.
	
	Regarding the Stokes $V$ profiles, Zeeman signatures are detected in all observations, featuring amplitudes of typically 0.1\%. Expressed in units of the unpolarised continuum level $I_{\rm c}$, the average noise levels of the Stokes~$V$ LSD signatures (dominated here by photon statistics) range from 2.3~10$^{-4}$ to 3.9~10$^{-4}$ per 1.8~\kms\ velocity bin - with a median value of 2.8~10$^{-4}$.
	%\begin{figure}
	%	\centering
	%	\includegraphics[scale=0.6]{lsds.eps}
	%	\caption[]{LSD circularly-polarised (Stokes~$V$, top/red curve) and unpolarised (Stokes~$I$, bottom/blue curve) profiles of TAP~26 collected on 2015~Jan.~22 (cycle 90.980). A clear Zeeman signature is detected in the LSD Stokes~$V$ profile with a complex shape, in conjunction with the unpolarised line profile. The mean polarisation profile is expanded by a factor of 10 and shifted upwards by 1.02 for display purposes.  }
	%	\label{fig:lsd}
	%\end{figure}
	
	The emission core of the \caii\ Infrared Triplet (IRT) lines exhibit an average equivalent width of $\simeq$10~\kms, corresponding to the amount expected from chromospheric emission for such a wTTS. The \hei\ $D_3$ line is relatively faint (average equivalent width of $\simeq$5~\kms), demonstrating that accretion is no longer taking place at its surface, in agreement with previous studies \citep{Donati14,Donati15}. The \hal\ line is also relatively weak by wTTS standards \citep{Kenyon95}, with an average equivalent width of 40~\kms, and is modulated with a period of 0.7132$\pm$0.0002~d (see App.~\ref{sec:act}).
	
	Contemporaneous VR$_{\rm J}$ photometric observations were also collected from the Crimean Astrophysical Observatory (CrAO) 1.25~m telescope between 2015~Aug and 2016~Mar. They indicate a brightness modulation with a period of 0.7138$\pm$0.0001~d of full amplitude 0.116~mag in \V\ (see Table~\ref{tab:pho}). By analogy with other wTTSs, these photometric variations can be safely attributed to the presence of brightness features at the surface of TAP~26 modulated by rotation. The small difference with the value found in \citet{Grankin13} suggests the presence of differential rotation in TAP~26 (see Sec.~\ref{sec:mod}).
	
	\section{Evolutionary status of TAP~26}
	\label{sec:evo}
	\begin{figure}
		\centering
		\includegraphics[scale=0.35,angle=-90]{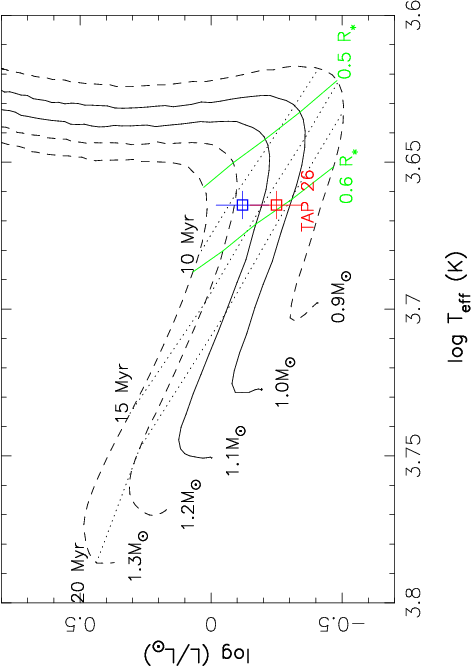}
		\caption[]{Observed location of TAP~26 in the HR diagram. The red and blue open squares (with 1$\sigma$ error bars) depict the location of TAP~26 using 2 different ways of estimating the inclination angle of the rotation axis - with the red one showing our best estimate used throughout the paper. The PMS evolutionary tracks for 0.9~\msun, 1.0~\msun, 1.1~\msun, 1.2~\msun\ and 1.3~\msun, and corresponding isochrones for 10~Myr, 15~Myr and 20~Myr \citep{Siess00} assume solar metallicity and include convective overshooting. The green lines depict where models predict PMS stars' radiative core reaches a radius of 0.5~\rstar\ and 0.6~\rstar.}
		\label{fig:hrd}
	\end{figure}
	TAP~26 is a well-studied single wTTSs, close enough to T~Tau, both spatially and in terms of velocity, to assume a distance of 147$\pm$3~pc \citep{Loinard07, Torres09}, with an error bar similar to that found on other regions of Taurus like L1495.
	
	Applying the automatic spectral classification tool especially developed in the context of MaPP (Magnetic Protostars and Planets) and MaTYSSE, following that of \citet{Valenti05} and discussed in \citet{Donati12}, we find that
	the photospheric temperature and logarithmic gravity of TAP~26 are respectively equal to \teff=4,620$\pm$50~K and \logg=4.5$\pm$0.2 (with $g$ in cgs units). This is warmer than the temperature quoted in the literature \citep[4340~K,][]{Grankin13}, which is derived from photometry and thus expected to be significantly less accurate than ours, derived from high-resolution spectroscopic data, enabling to find the actual temperature without the disturbance of circumstellar and interstellar reddening.
	
	Long-term photometric monitoring of TAP~26 indicates that its maximum \V\ magnitude is equal to 12.16 \citep{Grankin08}. Following \citet{Donati14,Donati15}, we assume a spot coverage\footnote{Spot coverage: integral of the difference between local brightness and photosphere brightness over the surface of the star, in units of photosphere brightness.} of $\simeq$25\% at maximum brightness, typical for active stars (and caused by, e.g., the presence of high-latitude cool spots and~/~or of small spots evenly spread over the whole stellar surface), we derive an unspotted \V\ magnitude of 11.86$\pm$0.20. From the difference between the \BmV\ index expected at the temperature of TAP~26 \citep[equal to 0.99$\pm$0.02,][]{Pecaut13} and the averaged value measured for TAP~26 \citep[equal to 1.13$\pm$0.05, see][]{Kenyon95, Grankin08}, and given the very weak impact of starspot on B-V \citep{Grankin08}, we derive that the amount of visual extinction \AV\ that our target suffers is equal to 0.43$\pm$0.15 \citep[within 1.5$\sigma$ of the value of][despite the very different methods used to estimate this parameter]{Herczeg14}. Using the visual bolometric correction expected for the adequate photospheric temperature \citep[equal to -0.55$\pm$0.05, see][]{Pecaut13} and the distance estimate assumed previously (147$\pm$3~pc), corresponding to a distance modulus of 5.84$\pm$0.04, we finally obtain a bolometric magnitude of 5.04$\pm$0.26, or equivalently a logarithmic luminosity relative to the Sun of -0.12$\pm$0.10. Coupling with the photospheric temperature obtained previously, we find a radius of 1.36$\pm$0.17~\rsun\ for our target star.
	\begin{table}
		\centering
		\begin{tabular}{|ccc|}
			\hline
			Parameter & Value & Reference \\
			\hline
			$d$ (pc) & 147$\pm$3 & T09 \\
			\mstar\ (\msun) & 1.04$\pm$0.10 & \\
			\rstar\ (\rsun) & 1.17$\pm$0.17 & \\
			\teff\ (K) & 4,620$\pm$50 & \\
			\logg & 4.5 & \\
			\logL & -0.25$\pm$0.10 & \\
			Age (Myr) & $\simeq$17 & \\
			\Prot\ (d) & 0.7135 & G13 \\
			$i$ (\degr) & 55$\pm$10 & \\
			\vsini\ (\kms) & 68.2$\pm$0.5 & \\
			\omeq\ (\rpd) & 8.8199$\pm$0.0003 & \\
			\dom\ (\rpd) & 0.0492$\pm$0.0010 & \\
			%Spottedness & 0.12 & \\
			\vrad\ (\kms) & 16.25$\pm$0.20 & \\ 
			\hline
		\end{tabular}
		\caption{Parameters for TAP~26, inferred from the photometric and spectroscopic measurements and the ZDI analysis (see Sec.~\ref{sec:mod}). Respectively: distance to Earth $d$, mass \mstar, radius \rstar, effective temperature \teff, decimal logarithm of surface gravity \logg, logarithmic luminosity \logL, age, rotation period \Prot, inclination of the rotation axis to the line of sight $i$, line-of-sight-projected equatorial rotation velocity \vsini, equatorial rotation rate \omeq, difference \dom\ between equatorial and polar rotation rates and mean radial velocity in the barycentric rest frame \vrad\ (which was derived from our spectropolarimetric runs, see Sec.~\ref{sec:mod}). T09 and G13 in the references respectively stand for \citet{Torres09} and \citet{Grankin13}.}
		\label{tab:sum}
	\end{table}
	
	The rotation period of TAP~26 is well determined from long-term multi-colour photometric monitoring, with an average value over the full data set equal to 0.7135~d \citep{Grankin13}. Coupling this rotation period along with our measurements of the line-of-sight-projected equatorial rotation velocity \vsini\ of TAP~26 (equal to 68.2$\pm$0.5~\kms, see Sec.~\ref{sec:mod}), we can infer that \rstar$\sin i$=0.96$\pm$0.05~\rsun, where \rstar\ and $i$ denote the radius of the star and the inclination of its rotation axis to the line of sight.  Comparing with the radius derived from the luminosity and photometric temperature, we derive that $i$=45$\pm$8\degr.
	\begin{figure*}
		\centering
		\includegraphics[scale=0.65,angle=-90]{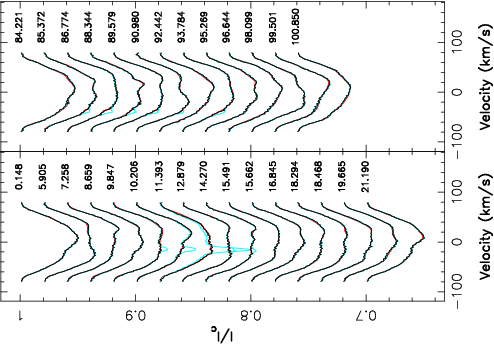}\hspace{3mm}
		\includegraphics[scale=0.65,angle=-90]{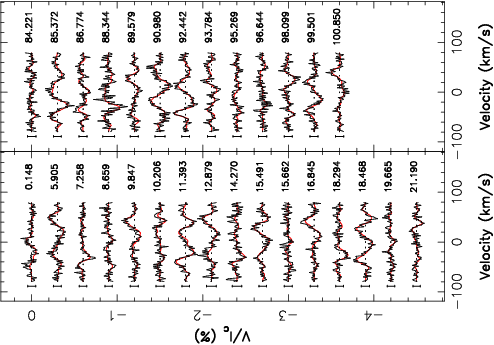}
		\caption[]{Maximum entropy fit (thin red lines) to the observed (thick black lines) Stokes~$I$ (left) and $V$ (right) LSD profiles. The 2015 Nov dataset is represented in the 1st and 3rd panels and the 2016 Jan dataset in the 2nd and 4th panels. The Stokes $I$ LSD profiles before the removal of lunar pollution are coloured in cyan, and 3$\sigma$-error bars are displayed for the Stokes $V$ profiles. The rotational cycles are written beside their corresponding profiles, in concordance with Table~\ref{tab:log}.}
		\label{fig:fit}
	\end{figure*}
	
	Using ZDI, we actually infer from our data that $i$=55$\pm$10\degr (see Sec.~\ref{sec:mod}). The 1$\sigma$ difference with the previous estimate can be simply interpreted as an over-estimate in spottedness at maximum brightness. Assuming now a spottedness of 12\% at maximum brightness (instead of 25\%) reconciles both approaches and yields a logarithmic luminosity of -0.25$\pm$0.10 and thus a radius of 1.17$\pm$0.17~\rsun, in good agreement with other studies \citep[1.18~\rsun\ in][]{Herczeg14}.
	
	Using the evolutionary models of \citet{Siess00} (assuming solar metallicity and including convective overshooting), we obtain that TAP~26 is a $\simeq$17~Myr star \citep[in good agreement with the estimate of][]{Grankin13} and that its mass is \mstar=1.04$\pm$0.10~\msun\ (see Fig~\ref{fig:hrd}). The average equivalent width of the 670.7~nm Li line is equal to 0.045~nm, in good agreement with that measured for solar-mass PMS stars in the 10-15~Myr Sco-Cen association at  the corresponding temperature \citep{Pecaut16}, which further confirms our age estimate and thus the evolutionary status of TAP~26.
	
	Referring to \citet{Donati15,Donati16a}, TAP~26 closely resembles an evolved version of the 2~Myr star V830~Tau that would have contracted and spun up by 4x towards the zero-age main sequence, with the rotation period and radius of V830~Tau being respectively 2.741~d and 2.0$\pm$0.2~\rsun. The increase in rotation rate matches quite well the predicted decrease in the moment of inertia between both epochs according to evolutionary models of \citet{Siess00}. Given the prominent role of the disc in braking the rotation of the star and thus decreasing its angular momentum \citep{Gallet15,Davies14}, this also suggests that TAP~26 dissipated its accretion disc very early, typically as early as, or earlier than V830~Tau. We also note that our target is located past the theoretical threshold at which stars start to be more than half radiative in radius, suggesting that the magnetic field of TAP~26 already started to evolve into a complex topology \citep{Gregory12}.
	
	The stellar parameters inferred and used in this study are summarised in Table~\ref{tab:sum}.
	\section{Tomographic imaging}
	\label{sec:mod}
	In order to model the activity jitter of TAP 26 (see Sec.~\ref{sec:mpl}), we applied ZDI \citep{Semel89, Brown91, Donati97c} to our data. ZDI takes inspiration from medical tomography, which consists of constraining a 3D distribution using series of 2D projections as seen from various angles \citep{Vogt87}. In our context, ZDI inverts simultaneous time series of 1D Stokes $I$ and $V$ LSD profiles into 2D brightness and magnetic field maps of the stellar surface \citep[see][]{Donati14}. The magnetic field is decomposed into its poloidal and toroidal components, both expressed as spherical harmonics expansions \citep{Donati06}.
	
	Synthetic LSD profiles are derived from brightness and magnetic maps by summing up the spectral contribution of all cells, taking into account the Doppler broadening caused by the rotation of the star, the Zeeman effect induced by magnetic fields and the continuum center-to-limb darkening. Local Stokes $I$ and $V$ profiles are computed using Unno-Rachkovsky's analytical solution to the polarised radiative transfer equations in a Milne-Eddington model atmosphere \citep{Landi04}. The local profile used in this study has a central wavelength, a Doppler width and a Land\'e factor of typical values 670~nm, 1.8~\kms\ and 1.2 respectively, and an equivalent width of 4.6~\kms\ corresponding to the LSD profiles of TAP~26. Technically, ZDI applies a conjugate gradient technique to iteratively reconstruct the brightness and magnetic surface maps with minimal information content (i.e. maximum Shannon entropy) that matches our observed LSD profiles at a given reduced chi-square (\chisqr, defined as $\chi^2$ divided by the number of data points\footnote{This follows the usual convention in regularised tomographic imaging techniques, where the number of model parameters is much smaller than the number of fitted data points and not taken into account in the expression of \chisqr\ \citep{Donati17}.}) level. Concerning the brightness, we note that, unlike in \citet{Donati97a} where we fit a spot filling factor with preset spot parameters, here we fit the local brightness $b_k$ of cell $k$, relative to the quiet photosphere (0$<b_k<$1 for dark spots and $b_k>$1 for bright plages), as described in \citet{Donati14}.
	
	ZDI can also take into account and model latitudinal differential rotation, shearing the brightness distribution and magnetic topology at the surface of the star, and assuming a solar-like surface rotation rate, $\Omega(\theta)$, varying with latitude, $\theta$, as:
	\begin{equation}
	\Omega(\theta)=\omeq-\dom (\sin\theta)^2
	\label{eqn:dr}
	\end{equation}
	where \omeq\ is the equatorial rotation rate and \dom\ is the difference between the equatorial and the polar rotation rates.
	
	For a given set of parameters, ZDI looks for the map with minimal information content that matches the LSD profiles at \chisqr=1. As a by-product, we obtain the optimal stellar parameters for which the reconstructed images contain minimal information: $i$=55$\pm$10\degr, \vsini=68.2$\pm$0.5~\kms\ and \vrad=17.0$\pm$0.2~\kms\ (the RV the star would have if unspotted and planet-free). Regarding differential rotation, we obtain \omeq=8.8199$\pm$0.0003~\rpd\ and \dom=0.0492$\pm$ 0.0010~\rpd, as outlined in more detail in Sec.~\ref{sec:par}.
	
	\subsection{Brightness and magnetic imaging}
	\label{sec:ima}
	Given the long time span between our two data sets (about 60~d, see Table~1), we start by reconstructing separate brightness and magnetic maps for each epoch (2015~Nov and 2016~Jan), before investigating the temporal variability between both in more detail.
	\begin{figure}
		\centering
		\includegraphics[scale=0.4,angle=-90]{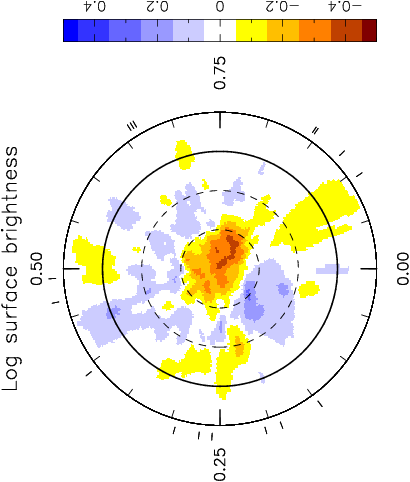}
		\includegraphics[scale=0.4,angle=-90]{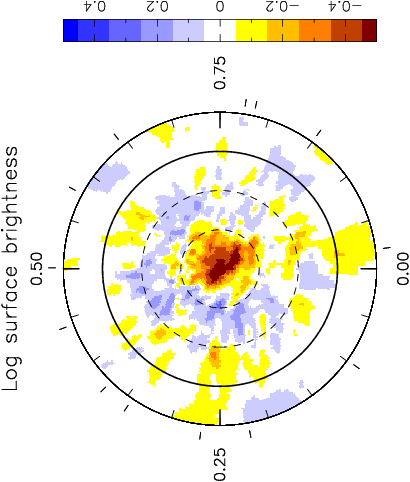}
		\caption[]{Flattened polar view of the surface brightness maps for the 2015 Nov dataset (top panel) and 2016 Jan dataset (bottom panel). The equator and the 60\degr, 30\degr\ and -30\degr\ latitude parallels are depicted as solid and dashed black lines respectively. The colour scale indicates the logarithm of the relative brightness, with brown/blue areas representing cool spots/bright plages. Finally, the outer ticks mark the phases of observation.}
		\label{fig:mapi}
	\end{figure}
	\begin{figure}
		\centering
		\includegraphics[scale=0.45]{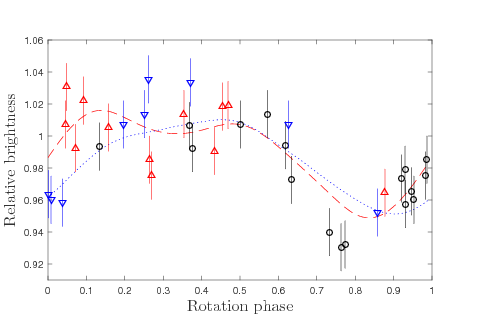}
		\caption{Photometry curves of the relative brightness as function of the rotation phase. The light curves synthesised from the reconstructed brightness maps for 2015 Nov and 2016 Jan are represented by a dashed red line and a dotted blue line respectively. The CrAO measurements are represented as dots with 1$\sigma$ error bars, with the observations from 2015 Aug to 2015 Oct in black circles, the observations from 2015 Oct to 2015 Dec in red upward-pointing triangles and the observations from 2015 Dec to 2016 Mar in blue downward-pointing triangles.}
		\label{fig:pho}
	\end{figure}
	
	The Stokes $I$ and $V$ LSD profiles, which are displayed in Fig.~\ref{fig:fit}, were used simultaneously to reconstruct both surface brightness and magnetic field maps. The synthetic LSD profiles presented in the figure match the observed ones at \chisqr=1, or, equivalently, at a $\chi^2$ equal to 1484 for the 2015 Nov dataset and 1157 for the 2016 Jan dataset, and for both sets of Stokes $I$ and $V$ LSD profiles. The iterative reconstruction starts from unspotted magnetic maps corresponding to \chisqr=13 (2015 Nov) and 9 (2016 Jan), showing that the iterative algorithm of ZDI successfully manages to reproduce the data at noise level. In the particular case of Stokes $I$ profiles, whose noise includes a significant level of systematics (see Sec.~\ref{sec:obs}), we find that smaller error bars make ZDI unable to fit the data down to \chisqr=1; on the opposite, greater error bars result in a fit to the Stokes $I$ profiles for which the raw radial velocities are not properly reproduced (see Sec.~\ref{sec:mpl}). This gives us confidence that the \sn\ values derived for the Stokes $I$ LSD profiles (see Table~\ref{tab:log}) are accurate and reliable within 10\%.
	\begin{figure*}
		\centering
		\includegraphics[scale=0.9,angle=-90]{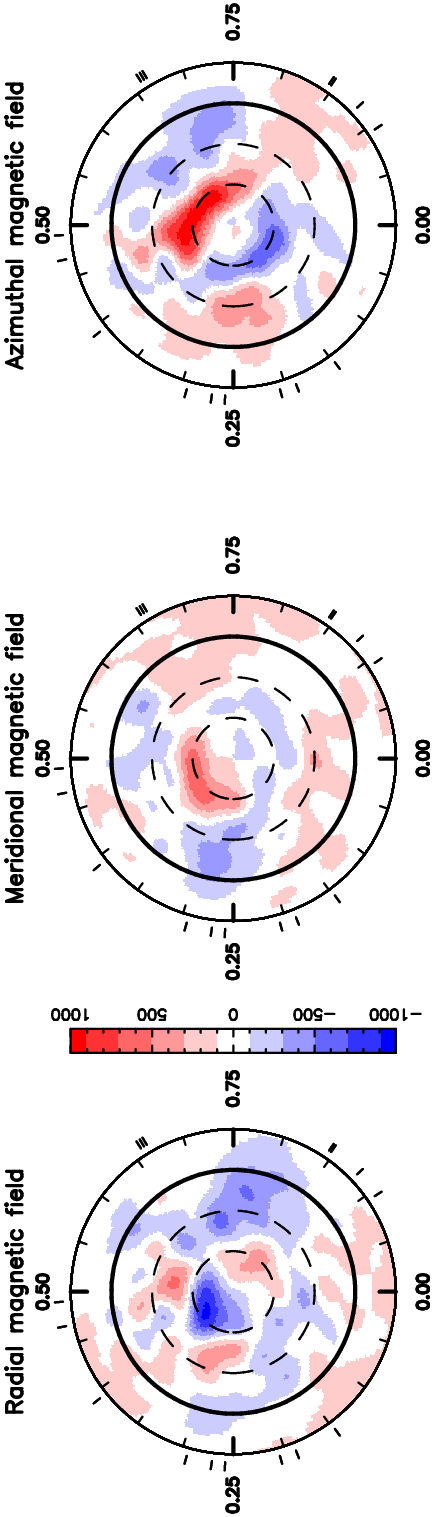}
		\includegraphics[scale=0.9,angle=-90]{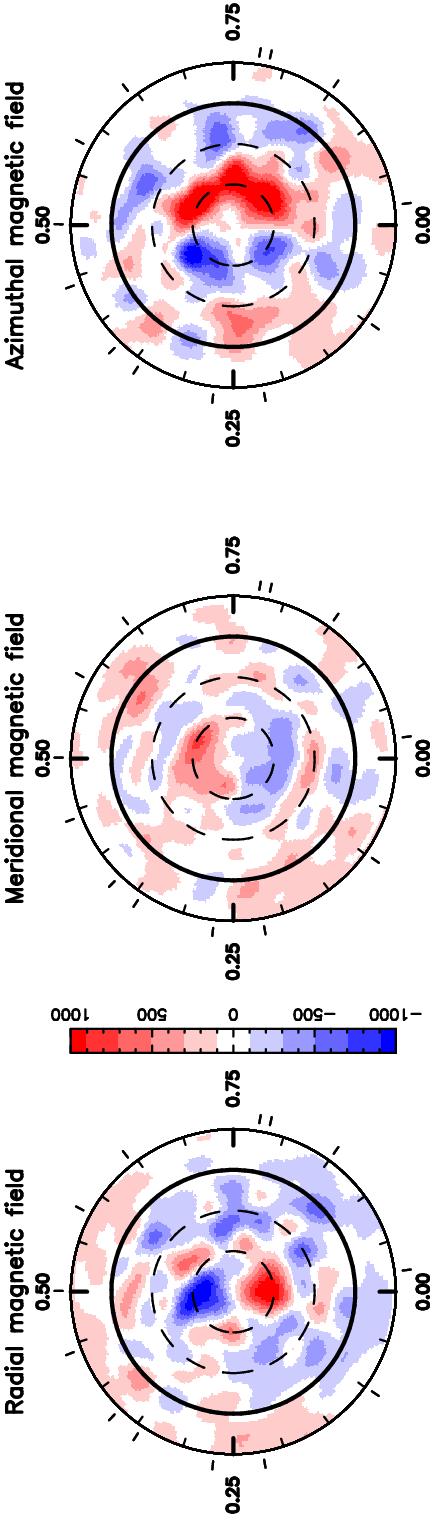}
		\caption[]{From left to right: radial, meridional and azimuthal component of the surface magnetic field (labelled in~G), reconstructed with ZDI from the 2015 Nov dataset (top panels) and the 2016 Jan dataset (bottom panels).}
		\label{fig:mapv}
	\end{figure*}
	\begin{figure*}
		\centering
		\includegraphics[scale=0.5,angle=-90]{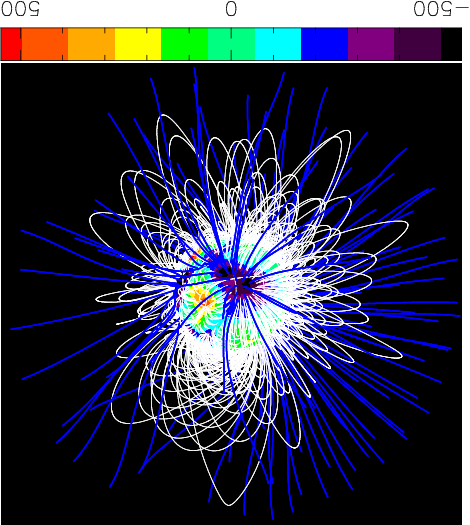}
		\includegraphics[scale=0.5,angle=-90]{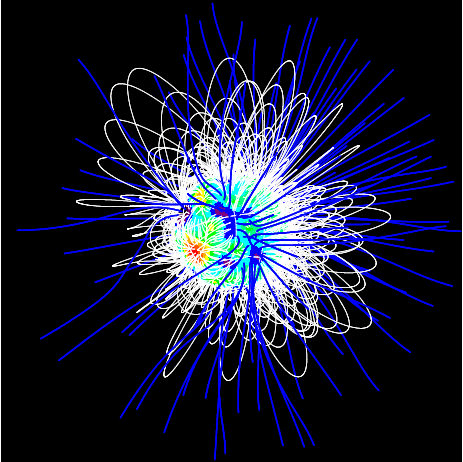}
		\caption[]{Potential field extrapolations of the reconstructed magnetic topology as seen by an Earth-based observer, in 2015 Nov (left) and in 2016 Jan (right) both at phase 0.8. Open and closed field lines are shown in blue and white respectively, whereas colours at the stellar surface depict the local value of the radial field (in G, as shown in the left panels of Fig.~\ref{fig:mapv}). The source surface at which the field becomes radial is set to 4~\rstar, slightly larger than the corotation radius of about 3~\rstar\ (at which the Keplerian period equals the stellar rotation period) and beyond which field lines are expected to quickly open under centrifugal forces.}
		\label{fig:lsf}
	\end{figure*}
	
	The reconstructed brightness maps for 2015 Nov and 2016 Jan are shown in Fig.~\ref{fig:mapi}, at an epoch corresponding to rotation cycle 10.0 (in the ephemeris of Eq.~\ref{eq:eph}) for 2015~Nov, and 92.0 for 2016~Jan (see Table~\ref{tab:log}); the colour scale codes the logarithmic relative brightness compared to that of the photosphere. The surface spot coverage we derive is similar at both epochs, reaching 10\% in the 2015 Nov map (5\% / 5\% of cool spots / hot plages respectively) and 12\% in the 2016 Jan map (7\% / 5\% of cool spots / hot plages respectively). Both reconstructed maps share some similarities, such as a large cool polar cap resembling that reconstructed on other rapidly rotating wTTSs \citep[e.g.][]{Skelly10, Donati14}, plus a number of smaller features located at lower latitudes (in particular the two equatorial spots located at phases 0.22 and 0.92 in 2015 Nov, 0.27 and 0.97 in 2016 Jan) interleaved with bright plages. We stress that ZDI is only sensitive to the medium and large brightness features and misses small spots evenly distributed over the whole stellar surface, implying that the spottedness we recover for TAP26 is likely an underestimate. We observe a number of differences between both images potentially attributable to differential rotation and / or intrinsic variability (see Sec.~\ref{sec:par}); however, the limited phase coverage at both epochs makes the direct comparison of individual surface features between maps ambiguous and hazardous. We caution that the smallest-scale structures may reflect to some extent the limited phase coverage and be subject to phase ghosting \citep[e.g.][]{Stout99}.
	
	Using the brightness maps reconstructed with ZDI, we can predict photometric light curves at both epochs, which are found to compare well with our contemporaneous CrAO observations (see Fig.~\ref{fig:pho}). Note the small but significant temporal evolution of the light-curve that we predict between the two epochs; this variability is however not obvious from the observed photometric data given their limited sampling and comparatively large error bars (rms 16 mmag).
	
	The reconstructed magnetic topology is shown in Fig.~\ref{fig:mapv}. The large-scale field reconstructed for TAP 26 features a rms magnetic flux of 330 and 430 G in 2015 Nov and 2016 Jan respectively. The field is found to be mainly poloidal (70\% of the reconstructed magnetic energy), though with a significant toroidal component (30\% of the reconstructed magnetic energy). It is also largely axisymmetric (50\% and 80\% of the poloidal and toroidal field energy respectively).
	
	The dipolar component of the large-scale field has a strength of 120$\pm$10~G at both epochs, corresponding to about 10\% of the reconstructed poloidal field energy, and is tilted at 40$\pm$5\degr\ to the line of sight, i.e., midway to the equator, towards phase 0.73$\pm$0.03 and 0.85$\pm$0.03 in 2015 Nov and 2016 Jan respectively. The increase in the phase towards which the dipole is tilted suggests that intermediate to high latitudes (at which the dipole poles are anchored) are rotating more slowly than average by 0.19\%, i.e., with a period of $\simeq$0.7148~d; this is confirmed by the fact that the line-of-sight projected (longitudinal) magnetic field \Bl\ \citep[proportional to the first moment of the Stokes $V$ profiles, e.g.,][and most sensitive to the low-order components of the large-scale field]{Donati97b} exhibit a recurrence timescale of 1.0014$\pm$0.003~\Prot\ (see App.~\ref{sec:act}), i.e., slightly longer than \Prot\ by a similar amount. Higher order terms in the spherical harmonics expansion describing the field (in particular the quadrupolar and octupolar modes) get stronger between 2015 Nov and 2016 Dec, with total magnetic energies increasing from 85\% to 93\% of the poloidal field.
	
	Finally, we show a large-scale extrapolation of the magnetic field (under the assumption of a potential field) in Fig.~\ref{fig:lsf}. Similarly to the brightness maps, the magnetic maps seem to point to a variation of the surface topology between late 2015 and early 2016, which is not explained by differential rotation alone, though the limited phase coverage calls for caution when comparing features between those maps.
	
	The magnetic maps suggest that the magnetic topology at the rotation pole underwent a $\simeq$0.1 phase shift between both dates.
	
	\subsection{Intrinsic variability and differential rotation}
	\label{sec:par}
	\begin{figure}
		\centering
		\includegraphics[scale=0.3]{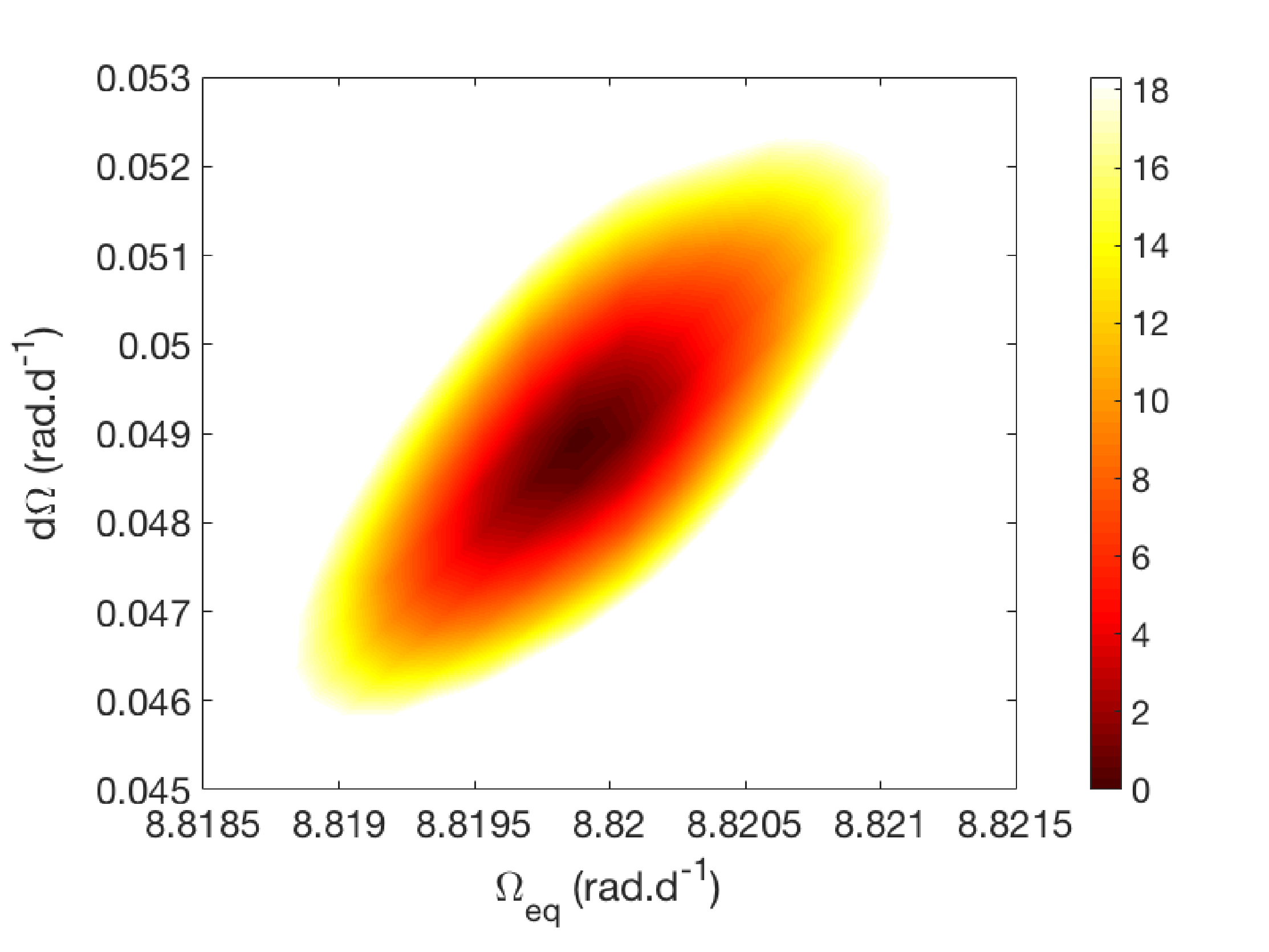}
		%\includegraphics[scale=0.6]{diffrotV.eps}
		%\caption[]{Variations of \chisqr\ as a function of \omeq\ and \dom, derived from the modelling of our Stokes~$I$ (upper) and $V$ (lower) LSD profiles of TAP~26 at constant information content. The full lines correspond to 68.3\%, 99\% and 99.99\% confidence level contours.}
		\caption[]{Map of \dchis\ as a function of \omeq\ and \dom, derived from the modelling of our Stokes~$I$ LSD profiles of TAP~26 at constant information content. A well-defined paraboloid is observed with the outer colour contour corresponding to the 99.99\% confidence level area (i.e., a $\chi^2$ increase of 18.4 for the 2581 Stokes $I$ data points). The minimum value of \chisqr\ is 1.4116. The minimum \chisqr\ achieved is above unity due to intrinsic variability affecting the LSD profiles but not being taken into account within ZDI. The derived differential rotation parameters are \omeq=8.8199$\pm$0.0003~\rpd\ and \dom=0.0492$\pm$0.0010~\rpd.}
		\label{fig:dr}
	\end{figure}
	\begin{figure*}
		\centering
		\includegraphics[scale=0.48]{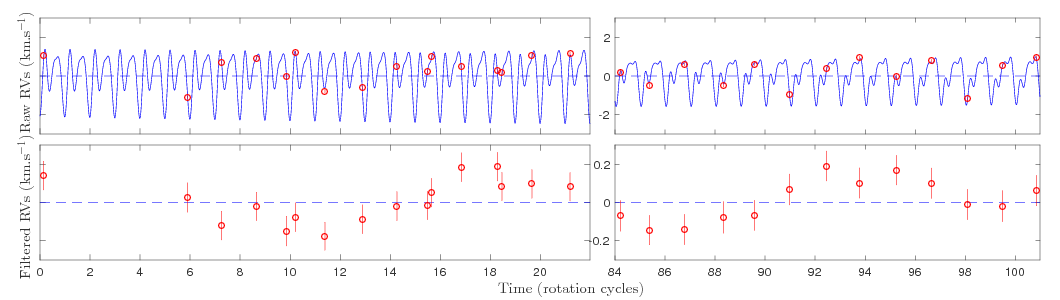}
		\caption[]{Top panels: RV (in the stellar rest frame) of TAP~26 as a function of rotation phase, as measured from our observations (open circles) and predicted by the tomographic maps (blue line). The synthesised raw RV curves exhibit only low-level temporal evolution resulting from differential rotation. Bottom panels: filtered RVs derived by subtracting the modelled activity jitter from the raw RVs, with a 10x zoom-in on the vertical axis.}
		\label{fig:rvs}
	\end{figure*}
	\begin{figure*}
		\centering
		\includegraphics[scale=0.6]{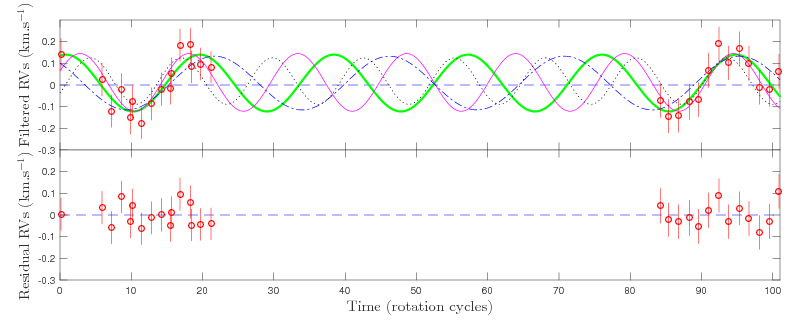}
		\caption[]{Top: filtered RVs of TAP~26 and four sine curves representing the best fits. The thick green curve represents the case \popr=18.80, the thin magenta one \popr=15.27, the dash-and-dotted blue one \popr=24.56 and the dotted black one \popr=12.76. Bottom: residual RVs resulting from the subtraction of the best fit (green curve) from the filtered RVs. The residual RVs feature a rms value of 51~\mps.}
		\label{fig:frv}
	\end{figure*}
	
	When applying ZDI to the whole dataset, i.e., modelling all Stokes $I$ and $V$ profiles with only one brightness map and one magnetic topology (see App.~\ref{sec:adf}), we obtain a minimum \chisqr\ value of 1.4, even when taking into account differential rotation (starting from an initial value \chisqr=20). This indicates that intrinsic variability occurred during the 45~d gap (or 63 rotation cycles) separating both datasets.
	
	Despite this variability, we attempted to retrieve differential rotation from the whole data set. The search for differential rotation parameters is done by minimising the value of \chisqr\ at a fixed amount of information, in this present case using the Stokes $I$ profiles and brightness map reconstruction only. From the curvature of the \chisqr\ paraboloid around the minimum, one can infer error bars on differential rotation parameters \citep{Donati03b}. The spot coverage is fixed at 13\% (chosen to be slightly higher than the values found in each reconstruction) and the values we found are \omeq=8.8199$\pm$0.0003~\rpd\ and \dom=0.0492$\pm$0.0010~\rpd, with a minimum \chisqr\ of 1.4116. A map of \dchis\ is shown in Fig.~\ref{fig:dr}, which presents a very clear paraboloid around the minimum we found, even if, due to our phase coverage, these precise values ask for further confirmation with the help of future data. This value of \dom\ is close to the solar differential rotation (0.055\rpd). The case with no differential rotation yields \chisqr=2.6907. Normalising $\Delta\chi^2$ by the minimum $\chi^2$ achieved over the map (to scale up error bars as a way to account for the contribution from the reported intrinsic variability) still yields a value in excess of 3300 and a negligible false alarm probability (FAP), unambiguously demonstrating that the star is not rotating as a solid body.
	
	The differential rotation parameters we obtain imply a lap time of 128$\pm$3~d, with rotation periods of 0.71239$\pm$0.00003~d and 0.71638$\pm$0.00008~d for the equator and pole respectively, in good agreement with the range of rotation periods derived from photometry \citep[ranging from 0.7135 to 0.7138,][]{Grankin13}. The 0.7132~d period found for the equivalent width of the \hal\ line and the 0.7145~d period found for the longitudinal magnetic field \Bl\ (see App.~\ref{sec:act}) are also consistent. We note that the rotation periods found with photometry, the longitudinal magnetic field and \hal\ line correspond to latitudes ranging from 30\degr\ to 50\degr, indicating that an important amount of activity is concentrated at these mid-latitudes, with the dipole pole located in the upper part of this range, in good agreement with the ZDI reconstruction (see Sec.~\ref{sec:ima}).
	
	%The differential rotation has been fitted using the Stokes $V$ LSD profiles and the magnetic topology reconstruction as well. Contrarily to the differential rotation deduced from the Stokes $I$ LSD profiles, there are several minima in the present case (Fig.~\ref{fig:dr}). The different minima might correspond to different scales of magnetic field features. The discrepancy in the values of \omeq\ and \dom\ indicate that the magnetic field is likely not embedded in the photosphere.
	\section{Modelling the planet signal}
	\label{sec:mpl}
	We describe below three different techniques aimed at characterising the RV curve of TAP~26. The first two are those used in \citet{Donati16a}: filtering out the activity modelled with the help of ZDI, and the simultaneous fit of the planet parameters and the stellar activity. The third method follows the approach of \citet{Haywood14} and \citet{Rajpaul15} and uses Gaussian-Process Regression (GPR) to model the activity directly from the raw RVs. The results obtained from these three methods are outlined and discussed in the following sections.
	\begin{figure*}
		\centering
		\includegraphics[angle=-90, scale=0.55]{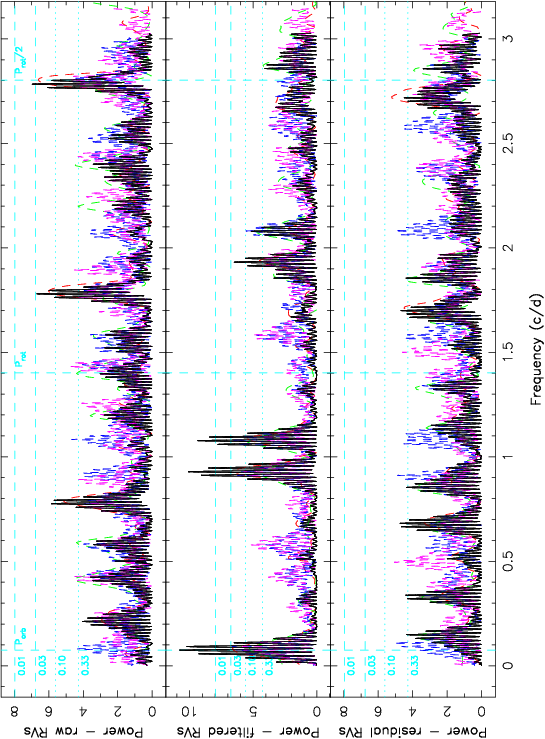}
		%\vspace{5mm}
		\includegraphics[angle=-90, scale=0.25]{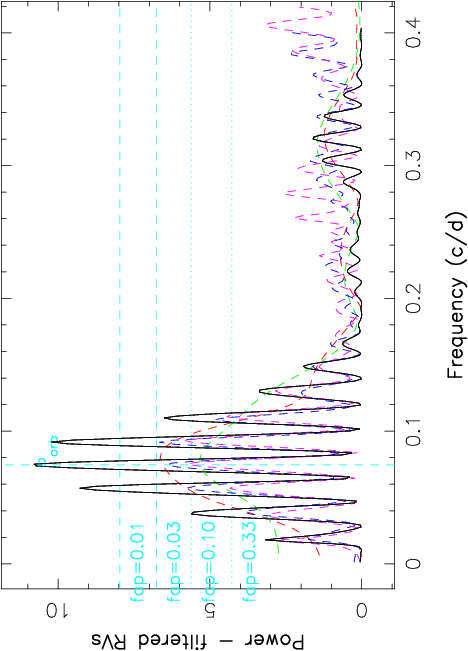}
		\vspace{-2mm}
		\caption{Top: Periodograms of the raw (top), filtered (middle) and residual (bottom) RV curves over the whole dataset (black line). The red line represents the 2015 Nov dataset, the green line the 2016 Jan dataset, the blue line the odd data points and the magenta line the even data points. FAP levels of 0.33, 0.10, 0.03 and 0.01 are displayed as horizontal dashed cyan lines. The rotation frequency (1.402 cycles per day) is marked by a cyan dashed line, as well as its first harmonic (2.803 cycles per day) and the frequency that has the smallest FAP (0.06\% at 0.075 cycles per day, corresponding to \Porb=13.41~d). Aliases of the highest peaks, related to the observation window, appear as lower peaks separated by one cycle per day. Bottom: Zoom in the periodogram of filtered RVs.}
		\label{fig:per}
	\end{figure*}
	\subsection{Jitter activity filtering}
	\label{sec:fil}
	\begin{table*}
		\centering
		\begin{tabular}{cccccccccc}
			\hline
			$K$ & \Porb & \Porb & $\phi$ & \BJDc & RV$_0$ & \chisqr & \dchis & $\log\mathcal{L}_{r1}$ & style \\
			(\kms) & (\Prot) & (d) & & (2,457,340+) & (\kms) & & & & in Fig.~\ref{fig:frv} \\
			\hline
			%		0.131       & 18.80      & 13.41      & 0.294       & 0.009       & 0.445 & 0 & 0.00 \\
			%		$\pm 0.013$ & $\pm 0.23$ & $\pm 0.16$ & $\pm 0.018$ & $\pm 0.010$ & & & \\
			%		\hline
			%		0.133       & 15.27      & 10.90      & 0.435       & 0.012       & 0.542 & 2.80 & -0.53 \\
			%		$\pm 0.015$ & $\pm 0.14$ & $\pm 0.10$ & $\pm 0.017$ & $\pm 0.011$ & & & \\
			%		\hline
			%		0.124       & 24.56      & 17.52      & 0.132       & 0.009       & 0.673 & 6.61 & -1.85 \\
			%		$\pm 0.017$ & $\pm 0.41$ & $\pm 0.30$ & $\pm 0.022$ & $\pm 0.013$ & & & \\
			%		\hline
			%		0.107       & 12.76      & 9.11       & 0.586       & 0.018       & 1.079 & 18.38 & -6.87 \\
			%		$\pm 0.021$ & $\pm 0.13$ & $\pm 0.09$ & $\pm 0.031$ & $\pm 0.015$ & & & \\
			0.131$\pm$0.020 & 18.80$\pm$0.23 & 13.41$\pm$0.16 & 0.709$\pm$0.026 & 8.75$\pm$0.35 & 0.009$\pm$0.014 & 0.445 & 0 & 0.00 & thick green \\
			\hline
			0.133$\pm$0.021 & 15.27$\pm$0.14 & 10.90$\pm$0.10 & 0.715$\pm$0.024 & 9.54$\pm$0.26 & 0.012$\pm$0.014 & 0.542 & 2.80 & -0.53 & thin magenta \\
			\hline
			0.124$\pm$0.020 & 24.56$\pm$0.41 & 17.52$\pm$0.30 & 0.684$\pm$0.028 & 7.11$\pm$0.50 & 0.009$\pm$0.016 & 0.673 & 6.61 & -1.85 & dashed blue \\
			\hline
			0.107$\pm$0.021 & 12.76$\pm$0.14 & 9.11$\pm$0.10 & 0.724$\pm$0.031 & 10.14$\pm$0.28 & 0.018$\pm$0.015 & 1.079 & 18.38 & -6.87 & dotted black \\
			\hline
			0 & & & & & 0.013$\pm$0.014 & 2.025 & 45.82 & -19.73 & \\
			\hline
		\end{tabular}
		\caption{Characteristics of the four best sine curve fits to the filtered RVs, and the case without planet. Respectively: semi-amplitude $K$, orbital period \Porb\ in units of \Prot, orbital period \Porb\ in days, phase of inferior conjunction $\phi$ relative to cycle 11.0 (see ephemeris in Eq.~\ref{eq:eph}), BJD of inferior conjunction, mean RV, corresponding \chisqr, difference in $\chi^2$ with the best fit (\dchis, summed on the 29 data points), and natural logarithm ($\log_e$) of the likelihood $\mathcal{L}_{r1}$ relative to the best fit. $\phi$ relates to the epoch of inferior conjuntion \BJDc\ through \BJDc=2,457,352.6485+$\phi$\Porb, the reference date being chosen so as to minimise the variation of $\phi$ between the four cases.}
		\label{tab:met1}
	\end{table*}
	The first technique consists of using the previously reconstructed maps to predict the pollution to the RV curve caused by activity (called activity jitter in the following) and subtract it from the raw RVs. From the observed Stokes $I$ LSD profiles, we compute, at both epochs, the raw RVs \rvraw\ (and error bars, see Table~\ref{tab:log}), as the first-order moment of the continuum-subtracted corresponding profiles \citep{Donati17}. Likewise, the synthesised Stokes $I$ LSD profiles derived from the brightness maps yield the synthesised activity jitter of the star (RV signal caused by the brightness distribution and stellar rotation). By subtracting the activity jitter from the raw RVs, we obtain filtered RVs \rvfil\ (see Table~\ref{tab:log}). We observe that the jitter has a mean semi-amplitude of 1.81~\kms\ in 2015 Nov and 1.21~\kms\ in 2016 Jan, whereas the filtered RV curve features a signal with a semi-amplitude of $\simeq$0.15~\kms\ (Fig.~\ref{fig:rvs}), i.e., 8 to 12 times smaller than the activity signal we filtered out. We note the very significant evolution in the activity curve between 2015 Nov and 2016 Jan, demonstrating that the brightness distribution has evolved at the surface of TAP~26, through differential rotation and intrinsic variability (see Sec.~\ref{sec:mod}).
	\begin{table*}
		\centering
		\begin{tabular}{cccccccc}
			\hline
			$K$ & \Porb & \Porb & $\phi$ & \BJDc & \chisqr & \dchis & $\log\mathcal{L}_{r2}$ \\
			(\kms) & (\Prot) & (d) & & (2,457,340+) & & & \\
			\hline
			%		0.153 & 15.30 & 10.92 & 0.387 & 0.96325 & 0.00 & 0.00 \\
			%		$\pm 0.024$ & $\pm 0.15$ & $\pm 0.11$ & $\pm 0.025$ & & & \\
			%		\hline
			%		0.143 & 18.81 & 13.42 & 0.266 & 0.96463 & 3.57 & -1.17 \\
			%		$\pm 0.023$ & $\pm 0.25$ & $\pm 0.18$ & $\pm 0.029$ & & & \\
			%		\hline
			%		0.145 & 12.84 & 9.16 & 0.531 & 0.96714 & 10.05 & -4.01\\
			%		$\pm 0.025$ & $\pm 0.12$ & $\pm 0.09$ & $\pm 0.028$ & & & \\
			%		\hline
			%		0.128 & 24.63 & 17.57 & 0.150 & 0.96888 & 14.54 & -6.10\\
			%		$\pm 0.026$ & $\pm 0.67$ & $\pm 0.48$ & $\pm 0.039$ & & & \\
			0.154$\pm$0.022 & 15.29$\pm$0.15 & 10.91$\pm$0.11 & 0.671$\pm$0.035 & 9.06$\pm$0.38 & 0.96824 & 0.00 & 0.00 \\
			\hline
			0.144$\pm$0.023 & 18.78$\pm$0.25 & 13.40$\pm$0.18 & 0.685$\pm$0.041 & 8.43$\pm$0.55 & 0.96979 & 4.00 & -1.34 \\
			\hline
			0.148$\pm$0.025 & 12.83$\pm$0.12 & 9.16$\pm$0.09 & 0.677$\pm$0.038 & 9.69$\pm$0.35 & 0.97180 & 9.17 & -3.61\\
			\hline
			%$0.128\pm 0.026$ & $24.63\pm 0.67$ & $17.57\pm 0.48$ & $0.150\pm 0.039$ & 0.96888 & 14.54 & -6.10\\
			%\hline
			0 & & & & & 0.98631 & 46.62 & -21.60 \\
			\hline
		\end{tabular}
		\caption{Optimal orbital parameters derived with the method described in Sec.~\ref{sec:pla}, respectively: semi-amplitude $K$, orbital period \Porb\ in units of \Prot, orbital period \Porb\ in days, phase of inferior conjunction $\phi$ relative to cycle 11.0, BJD of inferior conjunction, \chisqr, \dchis\ summed on 2581 data points, and natural logarithm of the likelihood $\mathcal{L}_{r2}$ relative to the best fit. The case where no planet is taken into account in the model is given for comparison.}
		\label{tab:met2}
	\end{table*}
	
	With a rms dispersion of 109~\mps, the filtered RVs clearly show the presence of a signal. Looking for a planet signature, we want to fit a sine curve (of semi-amplitude $K$, period \Porb, phase of inferior conjunction $\phi$, and offset RV$_0$) to these filtered RVs, which corresponds to a circular orbit (see Fig.~\ref{fig:frv}). The phase of inferior conjunction, i.e., corresponding to the epoch at which the planet is closest to us, is defined relatively to the reference date \BJDc$_0$=2,457,352.6485 (rotation cycle 11.0, approximately at the centre of the 2015 Nov observation run), such that the inferior conjunction occurs at \BJDc=\BJDc$_0$+($\phi$-1)\Porb. Due to the gap between both observing runs, several sine fits with different frequencies match the \rvfil\ as local minima of \chisqr. The four best fits are shown in Fig.~\ref{fig:frv} and their characteristics are given in Table~\ref{tab:met1}, with the value of the log likelihood as computed from the $\Delta\chi^2$ over these 29 RV data points. The residual RVs, derived from subtracting the best sine fit to the filtered RVs (shown in Fig.~\ref{fig:frv}), feature a rms value of 51~\mps.
	
	Plotting Lomb-Scargle periodograms for the raw RVs, filtered RVs and residual RVs further demonstrates the presence of a periodic signal in the filtered RVs (Fig.~\ref{fig:per}). The above-mentioned dominant periods are seen as peaks in the periodogram; periodograms of partial data (only the 2015 Nov dataset, only the 2016 Jan dataset, odd points and even points) are also shown, yielding peaks at the same frequencies albeit with a lower power. We highlight the fact that the highest peaks in the raw RVs correspond to the activity jitter and are located at \Prot/2 and its aliases, whereas little power concentrates at \Prot\ itself. A zoom-in of the filtered RV periodogram is also shown in Fig.~\ref{fig:per} (bottom panel). The FAP is 0.06\% for the highest peak (\Porb~=~13.41~d~=~18.80~\Prot), and no significant period stands out in the residual RVs after filtering out both the activity jitter and the planet signal corresponding to the highest peak. We carried out simulations to ensure that the detected peaks are not generated by the filtering process, see details in App.~\ref{sec:sim}. Study of other activity proxies shows that the detected orbital periods are not present in the activity signal either (App.~\ref{sec:act}).
	
	By fitting the filtered RVs with a Keplerian orbit rather than a circular orbit, we obtain an eccentricity of 0.16$\pm$ 0.15, indicating that there is no evidence for an eccentric orbit \citep[following the precepts of][]{Lucy71}. We can thus conclude that the orbit of TAP~26~b is likely close to circular, or no more than moderately eccentric.
	
	\subsection{Deriving the planetary parameters from the LSD profiles}
	\label{sec:pla}
	\begin{figure}
		\centering
		\includegraphics[scale=0.6]{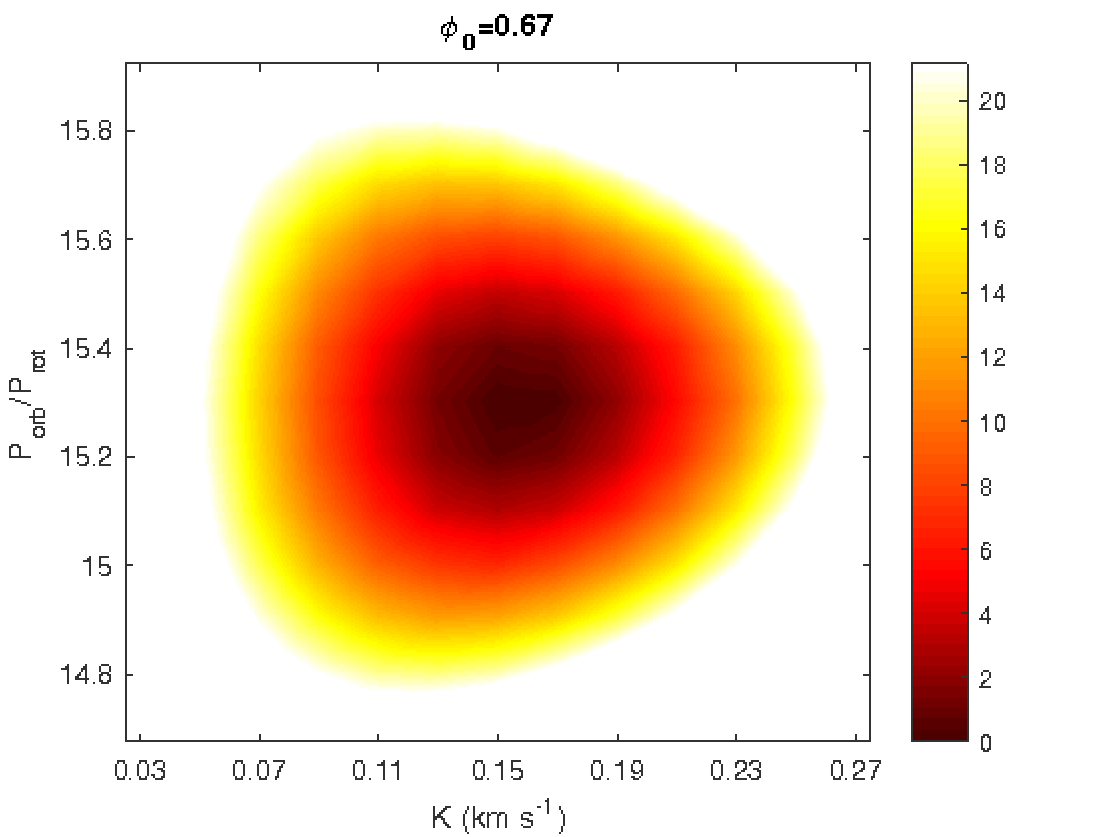}
		\caption[]{\dchis\ map as a function of $K$ and \popr, derived with ZDI from corrected Stokes $I$ LSD profiles at constant information content. Here the phase is fixed at 0.67, i.e., the value of $\phi$ at which the 3D paraboloid is minimum. The outer colour delimits the 99.99\% confidence level area (corresponding to a $\chi^2$ increase of 21.10 for 2581 data points in our Stokes $I$ LSD profiles). The minimum value of \chisqr\ is 0.96824.}
		\label{fig:mlp}
	\end{figure}
	A second technique, following the method of \citet{Petit15}, consists of taking into account the presence of a planet into the ZDI model. Rather than fitting the measured Stokes $I$ LSD profiles with a synthetic activity jitter directly, we first apply a translation in velocity to each of them, to remove the reflex motion caused by a planet of given parameters, and then apply ZDI to the corrected data set. Practically speaking, we repeat the experiment for a range of values for the orbital parameters ($K$, \Porb, $\phi$) at the vicinity of the minima previously identified in Sec.~\ref{sec:fil} and look for the set of values that yields the best result. The same way as for differential rotation, we derive the error bars on all parameters from the curvature of the 3D \chisqr\ paraboloid around the minimum.
	
	In the present case, since we have two datasets separated by a 45~d gap and we know that intrinsic variability occurred (see Sec.~\ref{sec:mod} and \ref{sec:fil}), a modification to the method described above was implemented: after correcting the global dataset from the reflex motion, ZDI is applied separately on each dataset, reconstructing two different brightness maps (one for late 2015 and one for early 2016) in order to obtain a more precise reconstruction. The quantity used to measure the likelihood of each set of parameters is therefore a global \chisqr, computed as a weighted average of both individual \chisqr, with respective weights proportional to the number of data points in each set (1424 for 2015 Nov and 1157 for 2016 Jan).
	
	As in the previous section, several minima are found, which are listed in Table~\ref{tab:met2}. We also computed the relative likelihood of each case compared to the best one from the corresponding difference in \chisqr. We note that the case with no planet yields \chisqr=0.98631, which leads to a relative probability lower than $10^{-9}$ compared to the case with a 10.91~d period planet.
	
	Figure~\ref{fig:mlp} displays a \dchis\ map around the local minimum \popr=15.29, at $\phi$=0.67, showing the 99.99\% confidence area.
	\begin{table*}
		\centering
		\begin{tabular}{ccccccc}
			\hline
			$K$ & \Porb & \Porb & $\phi$ & \BJDc & $\log\mathcal{L}$ & $\log\mathcal{L}_{r3}$ \\
			(\kms) & (\Prot) & (d) & & (2,457,340+) & & \\
			\hline
			0.163 & 12.61 & 8.99 & 0.766 & 10.54 & -3.48 & 0.00 \\
			$\pm$0.028 & $\pm$0.13 & $\pm$0.09 & $\pm$0.030 & $\pm$0.27 & & \\
			\hline
			0.149 & 15.12 & 10.79 & 0.728 & 9.71 & -3.73 & -0.25 \\
			$\pm$0.026 & $\pm$0.20 & $\pm$0.14 & $\pm$0.033 & $\pm$0.36 & & \\
			\hline
			0.139 & 18.74 & 13.37 & 0.694 & 8.56 & -5.60 & -2.12 \\
			$\pm$0.026 & $\pm$0.34 & $\pm$0.24 & $\pm$0.042 & $\pm$0.57 & & \\
			\hline
			0 & & & & & -15.80 & -12.52 \\
			\hline
		\end{tabular}
		\caption{Sets of orbital parameters that allow to fit the corrected RV curve best, using a GP with a covariance function given in Eq.~\ref{eqn:gp}, derived from the MCMC run. Respectively: reflex motion RV semi-amplitude $K$, orbital period \Porb\ in units of \Prot, orbital period \Porb\ in days, phase of inferior conjunction $\phi$ relative to rotation cycle 11.00 (ephemeris defined in Eq.~\ref{eq:eph}), BJD of inferior conjunction, natural logarithm of the marginal likelihood $\mathcal{L}$ and natural logarithm of the relative marginal likelihood $\mathcal{L}_{r3}$ as compared to the best case. The case where no planet is taken into account in the model is given for comparison.}
		\label{tab:gp}
	\end{table*}
	
	\subsection{Gaussian-Process Regression (GPR)}
	\label{sec:gau}
	The third method we used works directly with the raw RVs and aims at modelling the activity jitter and its temporal evolution with GPR, assuming it obeys an a priori covariance function \citep{Haywood14,Rajpaul15}. Similarly to the previous method, we fit both the orbit model and the jitter model simultaneously. For a planet with given parameters, we first remove the planet reflex motion from the RVs, then we fit the corrected RVs with a Gaussian process (GP) of pseudo-periodic covariance function:
	\begin{equation}
	c(t,t')=\theta_1^2.\exp\left[-\frac{(t-t')^2}{\theta_3^2}-\frac{\sin^2\left(\frac{\pi(t-t')}{\theta_2}\right)}{\theta_4^2}\right]
	\end{equation}
	where $t$ and $t'$ are two dates, $\theta_1$ is the amplitude (in \kms) of the GP, $\theta_2$ the recurrence timescale (in units of \Prot), $\theta_3$ the decay timescale (i.e., the typical spot lifetime in the present case, in units of \Prot) and $\theta_4$ a smoothing parameter (within [0,1]) setting the amount of high frequency structures that we allow the fit to include. From a given set of orbital parameters ($K$, \Porb, $\phi$) and of covariance function parameters ($\theta_1$ to $\theta_4$, called hyperparameters), we can derive the GP that best fits the corrected RVs (noted $y$ below) as well as the log likelihood $\log\mathcal{L}$ of the corresponding set of parameters from:
	\begin{equation}
	2\log\mathcal{L}=-n\log(2\pi)-\log |C+\Sigma |-y^T(C+\Sigma)^{-1}y
	\label{eqn:gp}
	\end{equation}
	where $n$ is the number of data points (29 in our case), $C$ is the covariance matrix of all the observing epochs and $\Sigma$ is the diagonal variance matrix of the raw RVs.
	\begin{figure*}
		\centering
		\hspace{-2mm}\includegraphics[scale=0.6,angle=-90]{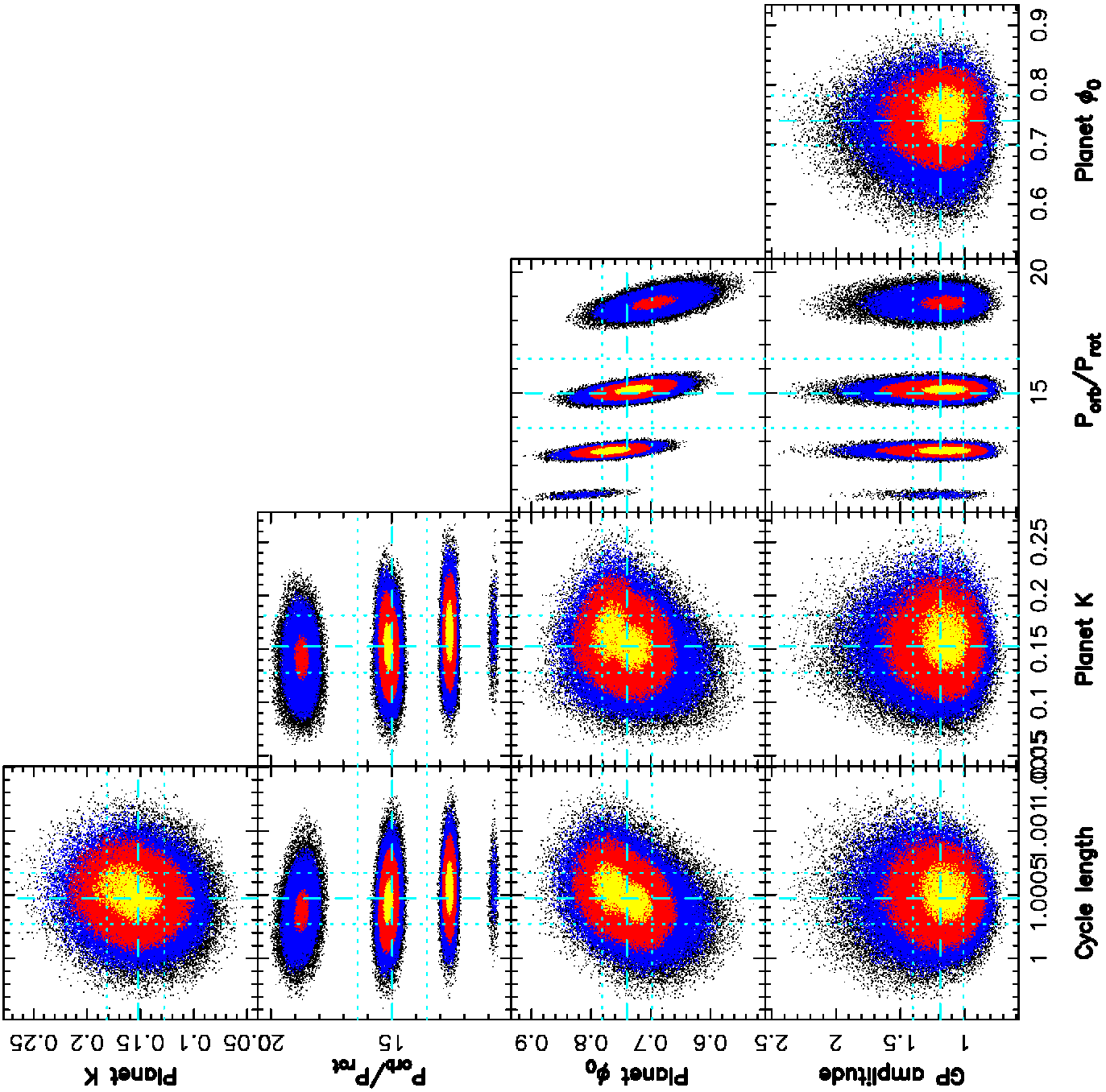}
		\caption{Phase plots of our 5-parameter MCMC run with yellow, red and blue points marking respectively the 1$\sigma$, 2$\sigma$ and 3$\sigma$ confidence regions. The optimal values found for each parameters are: $\theta_1$=1.19$\pm$0.21~\kms, $\theta_2$=1.0005$\pm$0.0002~\Prot, $K$=0.152$\pm$0.029~\kms. Several optima are detected for \Porb: 12.61$\pm$0.13~\Prot, 15.12$\pm$0.20~\Prot\ and 18.74$\pm$0.34~\Prot, ordered by decreasing likelihood. The corresponding phases $\phi$ are: 0.766$\pm$0.030, 0.728$\pm$0.033 and 0.694$\pm$0.042 respectively.}
		\label{fig:gpt}
	\end{figure*}
	\begin{figure*}
		\centering
		\includegraphics[scale=0.6,angle=-90]{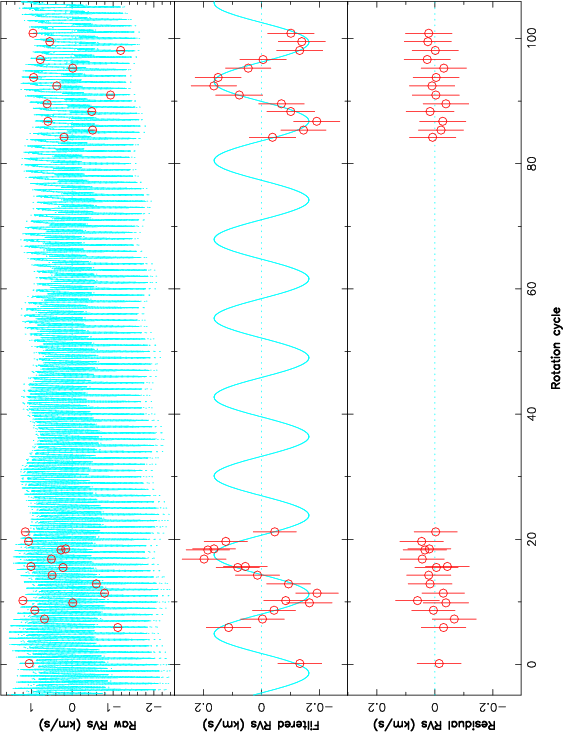}
		\caption[]{RV curves for a GPR fit of the activity jitter, with parameters $K$=0.163~\kms, \Porb=12.61~\Prot, $\phi$=0.766, $\theta_1$=1.19~\kms, $\theta_2$=1.0005~\Prot, $\theta_3$=180~\Prot, $\theta_4$=0.50~\Prot. Top panel: raw RVs and their error bars are shown in red, the solid cyan curve is the sum of the activity jitter predicted by GPR and the planet signal, and the dashed cyan lines show the 68.3\% confidence intervals about the prediction around this model. Middle panel: filtered RVs and their error bars, resulting from the subtraction of the GP-fitted activity jitter from the raw RVs (in red), and the sine curve corresponding to the assumed planet signal (in cyan). Bottom panel: residual RVs resulting from the subtraction of the planet signal from the filtered RVs, and their error bars. The residual RVs feature a rms value of 29~\mps, i.e. the GP fits the RVs down to \chisqr=0.151.}
		\label{fig:gps}
	\end{figure*}
	
	Coupled with a Markov Chain Monte Carlo (MCMC) simulation to explore the parameter domain, this method generates samples from the posterior probability distributions for the hyperparameters of the noise model and the orbital parameters. From these we can determine the maximum-likelihood values of these parameters and their uncertainty ranges. After an initial run where all the parameters are free to vary, we fix $\theta_4$ and $\theta_3$ to their respective best values (0.50$\pm$0.09 and 180$\pm$60~\Prot=128$\pm$43~d) before carrying out the main MCMC run to find the best estimates of the 5 remaining parameters. We note that the best value found for the decay time is exactly equal to the differential rotation lap time within error bars, and to twice the total span of our data. This decay time corresponds to both the differential rotation lap time and the starspot coherence time, since these are the most influent phenomena on the periodicity of the activity jitter. Such a starspot coherence time is consistent with previous studies \citep{Grankin08,Bradshaw14,Lanza06}.
	\begin{table*}
		\begin{tabular}{cccc}
			\hline
			& ZDI~\#1 & ZDI~\#2 & GPR \\
			\hline
			$K$ (\kms) & 0.133$\pm$0.021 & \textbf{0.154$\pm$0.022} & 0.149$\pm$0.026 \\
			$\phi$ & 0.715$\pm$0.024 & \textbf{0.671$\pm$0.035} & 0.728$\pm$0.033 \\
			\Porb\ (\Prot) & 15.27$\pm$0.14 & \textbf{15.29$\pm$0.15} & 15.12$\pm$0.20 \\
			\Porb\ (d) & 10.90$\pm$0.10 & \textbf{10.91$\pm$0.11} & 10.79$\pm$0.14 \\
			$a$ (au) & 0.0974$\pm$0.0032 & \textbf{0.0975$\pm$0.0032} & 0.0968$\pm$0.0032 \\
			\msini\ (\mjup) & 1.49$\pm$0.25 & \textbf{1.73$\pm$0.27} & 1.66$\pm$0.31 \\
			\BJDc\ (2,457,340+) & 9.54$\pm$0.26 & \textbf{9.06$\pm$0.38} & 9.71$\pm$0.36 \\
			\logLr & -0.53 & \textbf{0.00} & -0.25 \\
			$\theta_1$ (\kms) & & & 1.19$\pm$0.21 \\
			$\theta_2$ (\Prot) & & & 1.0004$\pm$0.0002 \\
			\hline
			$K$ (\kms) & 0.107$\pm$0.021 & 0.148$\pm$0.025 & \textbf{0.163$\pm$0.028} \\
			$\phi$ & 0.724$\pm$0.031 & 0.677$\pm$0.038 & \textbf{0.766$\pm$0.030} \\
			\Porb\ (\Prot) & 12.76$\pm$0.14 & 12.83$\pm$0.12 & \textbf{12.61$\pm$0.13} \\
			\Porb\ (d) & 9.11$\pm$0.10 & 9.16$\pm$0.09 & \textbf{8.99$\pm$0.09} \\
			$a$ (au) & 0.0864$\pm$0.0028 & 0.0868$\pm$0.0028 & \textbf{0.0858$\pm$0.0028} \\
			\msini (\mjup) & 1.13$\pm$0.23 & 1.56$\pm$0.28 & \textbf{1.71$\pm$0.31} \\
			\BJDc\ (2,457,340+) & 10.14$\pm$0.28 & 9.69$\pm$0.35 & \textbf{10.54$\pm$0.27} \\
			\logLr & -6.87 & -3.61 & \textbf{0.00} \\
			$\theta_1$ (\kms) & & & \textbf{1.19$\pm$0.21} \\
			$\theta_2$ (\Prot) & & & \textbf{1.0005$\pm$0.0002} \\
			\hline
		\end{tabular}
		\caption{Results yielded by the methods ZDI~\#1 (Sec.~\ref{sec:fil}), ZDI~\#2 (Sec.~\ref{sec:pla}) and GPR (Sec.~\ref{sec:gau}), for the two periods $\simeq$15~\Prot\ and $\simeq$13~\Prot. From top to bottom: reflex motion semi-amplitude $K$, phase of inferior conjunction $\phi$ relative to cycle 11.0, orbital period \Porb\ in units of \Prot, orbital period \Porb\ in days, semi-major axis $a$, \msini, BJD of inferior conjunction \BJDc, natural logarithm of relative likelihood as compared to the best case $\mathcal{L}_{r}$, GP amplitude $\theta_1$ and GP recurrence timescale $\theta_2$. Results are displayed in bold font when the period is found with the highest likelihood using the corresponding method.}
		\label{tab:sup}
	\end{table*}
	
	As shown in Fig.~\ref{fig:gpt}, this method succesfully recovers the different minima previously found with the first two techniques, with little correlation between the various parameters thus minimum bias in the derived values. Applying the method of \citet{Chib01} to the MCMC posterior samples, we obtain that the marginal likelihood of the case \Porb=12.61~\Prot\ is larger than that of the case \Porb=15.12~\Prot\ by a Bayes factor of only 1.28, which implies that there is as yet no clear evidence in favor of either of them. The third most likely case, \Porb=18.74~\Prot, has a marginal likelihood which is inferior to the first one by a Bayes' factor of $>8$, and the case with no planet has a marginal likelihood which is smaller than that of the first case by a Bayes factor of $2\ 10^5$. The three most likely sets of parameters are summarised in Table~\ref{tab:gp}.
	
	Trying to fit a non-circular Keplerian orbit to our data, i.e. adding the periapsis argument and the eccentricity $e$ to the parameters in our MCMC run, we obtain $e$=0.05$\pm$0.18, with a marginal likelihood slightly smaller than that of the case of a circular orbit. This further supports that the planet eccentricity is low if non-zero.
	
	The best fit with our third method is shown in Fig.~\ref{fig:gps}, where we see the raw RVs and the modelled RV curve predicted with this method, i.e., the sum of the GPR-fitted activity jitter and of the planet signal. Zooming in shows that this curve presents similarities with the RV jitter curve derived by ZDI (Fig.~\ref{fig:rvs}), indicating that, although working only with the RV data points, GPR successfully retrieves a convincing model for the activity. We also note the ability of the GP to model the activity jitter not only during our observing runs, but also during the 45~d gap between them, emphasising the variability of the RV signal with time. The residual RVs in the case presented here have a rms value of 29~\mps\ (close to the instrument RV precision 20-30~\mps) whereas the residual RVs derived with the 1st method yield a rms value of 51~\mps. Though the rms value is 2.5 times smaller than the error bar, GPR only fits 2 parameters, which illustrates its flexibility without decreasing its reliability, since the results are consistent with those found using independent methods (Sec.~\ref{sec:fil}, \ref{sec:pla}). This demonstrates that GPR does a better job at modelling the activity jitter and its temporal evolution than the 2 previous methods, in agreement with the conclusions of \citet{Donati17} in the case of the wTTS V830~Tau. As a result, we consider the optimal planet parameters derived with GPR as the most reliable ones, and therefore conclude that the orbital periods of 10.8 and 9.0~d are more or less equally likely.
	
	Table \ref{tab:sup} summarises the likelihood of the different periods found with each method.
	
	\section{Summary \& discussion}
	\label{sec:dis}
	This paper reports the results of an extended spectropolarimetric run on the wTTS TAP~26, carried out within the framework of the international MaTYSSE Large Programme, using the echelle spectropolarimeter ESPaDOnS at CFHT, spanning 72~d from 2015 Nov 18 to 2015 Dec 03 then from 2016 Jan 17 2016 Jan 29, and complemented by contemporaneous photometric observations from the 1.25~m telescope at CrAO.
	
	Applying Zeeman-Doppler Imaging (ZDI) to our two data sets, we derived the surface brightness and magnetic maps of TAP~26, revealing the presence of cool spots and warm plages totalling up to 12\% of the stellar surface (we however caution that this is a lower limit given the insensitivity of ZDI to small spots evenly spread over the stellar surface). The large-scale field of TAP~26 is found to be mainly poloidal and axisymmetric, with a 120~G dipole component tilted at 40\degr\ from the rotation axis. The 2015 Nov and 2016 Jan maps are mostly similar, but nonetheless feature some differences that indicate temporal evolution of the surface brightness and the magnetic field, demonstrated by the inability of ZDI to model the whole dataset at noise level, on a timescale comparable to that spanning our sample (72~d). ZDI also enabled us to detect the differential rotation pattern at the surface of TAP~26, with \dom=0.0492$\pm$0.0010~\rpd, a value close to that of the Sun, implying a time for the equator to lap the pole by one rotation equal to 128$\pm$3~d. .
	
	We then applied three different methods to search for a planetary signature in the observed spectra. The first method studies the radial velocities filtered out from the activity jitter predicted by ZDI. Our second method looks for the planet parameters that enable the best fit to the corrected LSD profiles, in a way similar to that used to estimate surface differential rotation. The third method uses Gaussian-Process Regression (GPR) to fit the activity jitter in the raw RVs, and like the second method, searches for the orbital parameters which enable GPR to fit the raw RVs corrected from the reflex motion best. We find that GPR succeeds best at modelling the intrinsic variability occurring at the surface of TAP~26, and is able to fit raw RVs at a rms precision of 29~\mps, i.e., close to the instrumental precision of ESPaDOnS \citep[20-30~\mps,][]{Moutou07,Donati08b} and 30\% better than with our first method (yielding a rms precision of 51~\mps). A similarly low rms was reached by GPR in the study of wTTS V830~Tau \citep[35-37~\mps,][]{Donati17}.
	
	All three methods demonstrate the clear presence of a planet signature in the data, although the gap between both data sets generates aliasing problems, causing multiple nearby peaks to stand out in the periodogram. Of the dominant periods, the 10.8~d one emerges strongly for all three methods. It is the most likely with the second method, and equally likely as other periods when using the first and third methods (13.4~d and 9.0~d respectively). Although the 9.0~d orbital period ranks low (and in particular lower than the 13.4~d period) with our first and second methods, we nonetheless consider it as the second most likely given its first rank with GPR; the most probable explanation for this apparent discrepancy lies in the higher ability of GPR at modelling intrinsic variability of the activity jitter plaguing the RV curve. Allowing ZDI to model temporal evolution of spot distributions and magnetic topologies should bring all methods on an equal footing; this upgrade is planned for a forthcoming study.
	
	Assuming the 10.79$\pm$0.14~d period is the true orbital period, and using the values yielded by GPR for $K$ and $\phi$, we find a circular orbit of semi-major axis $a$~=~0.0968~$\pm$~0.0032~au~=~17.8~$\pm$~2.7~\rstar, epoch of inferior conjunction \BJDc=2,457,349.71$\pm$0.36 and \msini=1.66$\pm$0.31~\mjup. If the orbital plane is aligned with the equatorial plane of TAP~26, with an assumed inclination of 55\degr, we obtain a mass $M$=2.03$\pm$0.46~\mjup\ for TAP~26~b. The 8.99$\pm$0.09~d period leads to $a$=0.086$\pm$0.003~au, \BJDc=2,457,350.54$\pm$0.27 and \msini=1.71$\pm$0.31~\mjup.
	
	With an age of $\simeq$17~Myr, TAP~26 is already an aging T~Tauri star and on the verge of becoming a post T~Tauri star, as demonstrated by its complex geometry and weaker dipole field component (consistent with TAP~26 having a mostly radiative interior). Akin to V830~Tau~b \citep{Donati16a}, the hJ in a nearly circular orbit that we have discovered in the young system TAP~26 is better explained by type II disc migration than by planet-planet scattering coupled to tidal circularisation. When compared to V830~Tau, a 2~Myr wTTS of similar mass \citep{Donati15, Donati16a, Donati17}, appears as an evolved version, rotating 4x faster than its younger sister, likely as a direct consequence of its 4x smaller moment of inertia \citep[according to the evolutionary models of][]{Siess00}.
	
	Regarding the hJs we detected around TAP~26 and V830~Tau and despite their differences (in mass in particular), it would be tempting to claim that, like its host star, TAP~26~b is an evolved version of V830~Tau~b. This would actually imply that TAP~26~b migrated outwards under tidal forces from a distance of $\simeq$0.057~au (where V830~Tau~b is located) to its current orbital distance of 0.094~au, as a result of the spin period of TAP~26 being $\simeq$15x shorter than the orbital period of TAP~26~b. This option seems however unlikely given the latest predictions of tidal interactions between a young T~Tauri star and its close-in hot Jupiter \citep{Bolmont16}, indicating that tidal forces can only have a significant impact on a hJ within 0.06~au of a solar-mass host star (for a typical TTS with a radius of $\simeq$2~\rsun). The most likely explanation we see is thus that TAP~26~b:  
	\begin{itemize}
		\item ended up its type-II migration in the accretion disc at the current orbital distance, when TAP~26 was still young, fully convective and hosting a large-scale dipole field of a few kG similar to that of AA~Tau \citep{Donati10b}, i.e., strong enough to disrupt the disc up to a distance of 0.09~au,
		\item was left over once the disc has dissipated at an age significantly smaller than 2~Myr, i.e., before the large-scale field had time to evolve into a weaker and more complex topology, and the inner accretion disc to creep in as a result of the decreasing large-scale field and the subsequent chaotic accretion \citep[e.g.,][]{Blinova16}.
	\end{itemize}
	Admittedly, this scenario requires favorable conditions to operate; in particular, it needs the accretion disc to vanish in less than 2~Myr, which happens to occur in no more than 10\% of single T~Tauri stars in Taurus \citep{Kraus12}. In fact, since both TAP~26 and V830~Tau have the same angular momentum content, it is quite likely that TAP~26 indeed dissipated its disc very early (see Sec.~\ref{sec:evo}). Quantitatively speaking, assuming (i) that the hJ we detected tracks the location of the inner disc when the disc dissipated, (ii) that the spin period at this time was locked on the Keplerian period of the inner disc (equal to the orbital period of the detected hJ) and (iii) that stellar angular momentum was conserved since then, we derive that the disc must have dissipated when TAP~26 was about three times larger in radius, at an age of less than 1~Myr \citep[according to][]{Siess00}. Generating a magnetospheric cavity of the adequate size (0.085 to 0.097~au depending on the orbital period) would have required TAP~26 to host at this time a large scale dipole field of 0.3-1.0~kG for mass accretion rates in the range ~10$^{-9}$ to 10$^{-8}$ \msun /yr, compatible with the large-scale fields found in cTTSs of similar masses \citep[e.g., GQ~Lup,][]{Donati12}.
	
	Along with other recent reports of close-in giant planets (or planet candidates) detected (or claimed) around young stars \citep{Eyken12,Donati16a,Donati17,Mann16,JohnsKrull16,David16}, our result may suggest a surprisingly high frequency of hJs around young solar-type stars, with respect to that around more evolved stars \citep[$\simeq$1\%,][]{Wright12}. However, this may actually reflect no more than a selection bias in the observation samples (as for their mature equivalents in the early times of velocimetric planet detections). Planets are obviously much easier to detect around non-accreting TTSs as a result of their lower level of intrinsic variability; observation samples (like that of MaTYSSE) are thus naturally driven towards young TTSs whose accretion discs vanished early, i.e., at a time when their large-scale fields were still strong and their magnetospheric gaps large, and thus for which hJs had more chances to survive type-II migration.  A more definite conclusion must wait for a complete analysis of the full MaTYSSE sample.
	
	More observations of TAP~26, featuring in particular a more regular temporal sampling, are currently being planned to better determine the characteristics of the newborn hJ we detected. Furthermore, analysing thoroughly the full MaTYSSE data set to pin down the frequency of newborn hJs within the sample observed so far will bring a clearer view on how the formation and migration of young giant planets is occurring. Ultimately, only a full-scale planet survey of young TTSs such as that to be carried out with SPIRou, the new generation spectropolarimeter currently being built for CFHT and scheduled for first light in 2018, will be able to bring a consistent picture of how young close-in planets form and migrate, how their population relates to that of mature hJs, and more generally how young hJs impact the formation and early architecture of planetary systems like our Solar System.
	
	\section*{Acknowledgements}
	
	This paper is based on observations obtained at the Canada-France-Hawaii Telescope (CFHT), operated by the National Research Council of Canada, the Institut National des Sciences de l'Univers of the Centre National de la Recherche Scientifique (INSU/CNRS) of France and the University of Hawaii.  We thank the CFHT QSO team for the great work and effort at collecting the high-quality MaTYSSE data presented in this paper. MaTYSSE is an international collaborative research programme involving experts from more than 10 different countries (France, Canada, Brazil, Taiwan, UK, Russia, Chile, USA, Ireland, Switzerland, Portugal, China and Italy). We also warmly thank the IDEX initiative at Universit\'e F\'ed\'erale Toulouse Midi-Pyr\'en\'ees (UFTMiP) for funding the STEPS collaboration program between IRAP/OMP and ESO. We acknowledge funding from the LabEx OSUG@2020 that allowed purchasing the ProLine PL230 CCD imaging system installed on the 1.25-m telescope at CrAO. SGG acknowledges support from the Science \& Technology Facilities Council (STFC) via an Ernest Rutherford Fellowship [ST/J003255/1]. SHPA acknowledges financial support from CNPq, CAPES and Fapemig.
	
	%%%%%%%%%%%%%%%%%%%%%%%%%%%%%%%%%%%%%%%%%%%%%%%%%%
	
	
	
	\bibliographystyle{mnras}
	\bibliography{tap26} % if your bibtex file is called example.bib
	
	%%%%%%%%%%%%%%%%%%%%%%%%%%%%%%%%%%%%%%%%%%%%%%%%%%
	
	%%%%%%%%%%%%%%%%% APPENDICES %%%%%%%%%%%%%%%%%%%%%
	
	\appendix
	
	\section{Additional figures}
	\label{sec:adf}
	\begin{figure}
		\centering
		\includegraphics[scale=1.35,angle=-90]{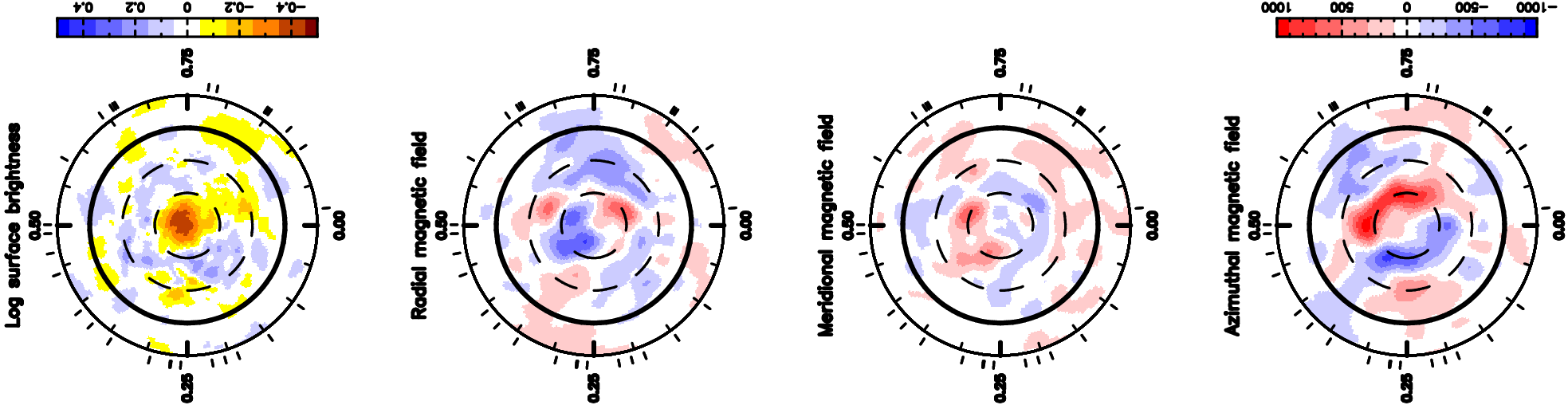}
		%	\caption[]{Brightness map when fitting the 2015 Nov and 2016 Jan datasets altogether, at rotation cycle 51.}
		\caption[]{Brightness and magnetic components surface maps when fitting the 2015 Nov and 2016 Jan datasets altogether, at rotation cycle 51.}
		\label{fig:maptQ}
	\end{figure}
	%\begin{figure*}
	%	\centering
	%	\includegraphics[scale=0.9,angle=-90]{novjanB_dr_B.ps}
	%	\caption[]{Radial, meridional and azimuthal magnetic field maps when fitting the 2015 Nov and 2016 Jan datasets altogether, at rotation cycle 51.}
	%	\label{fig:maptB}
	%\end{figure*}
	
	Images of brightness and magnetic field on the surface of TAP~26, as derived with ZDI using our 29 spectra, are shown in Fig.~\ref{fig:maptQ}.% and \ref{fig:maptB}.
	
	\section{Activity proxies}
	\label{sec:act}
	\begin{figure*}
		\centering
		\includegraphics[scale=0.55,angle=-90]{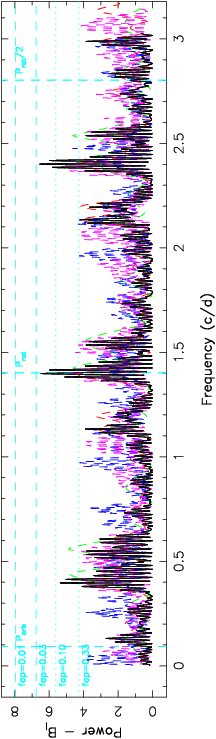}
		\caption[]{Periodogram of the longitudinal magnetic field. The rotation period at 0.7135~d is represented by a dashed vertical cyan line, as well as the orbital period at 10.92~d.}
		\label{fig:pbl}
	\end{figure*}
	\begin{figure*}
		\centering
		\includegraphics[scale=0.55,angle=-90]{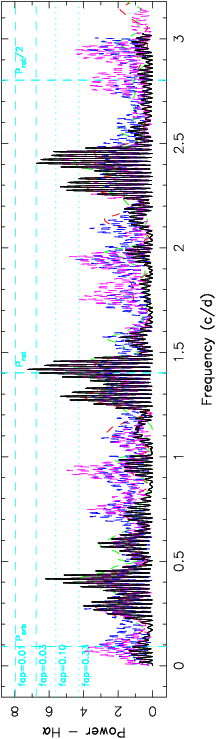}
		\caption[]{Periodogram of the \hal\ line equivalent width. The rotation period at 0.7135~d is represented by a dashed vertical cyan line, as well as the orbital period at 10.92~d.}
		\label{fig:pha}
	\end{figure*}
	\begin{figure}
		\centering
		\includegraphics[scale=0.9]{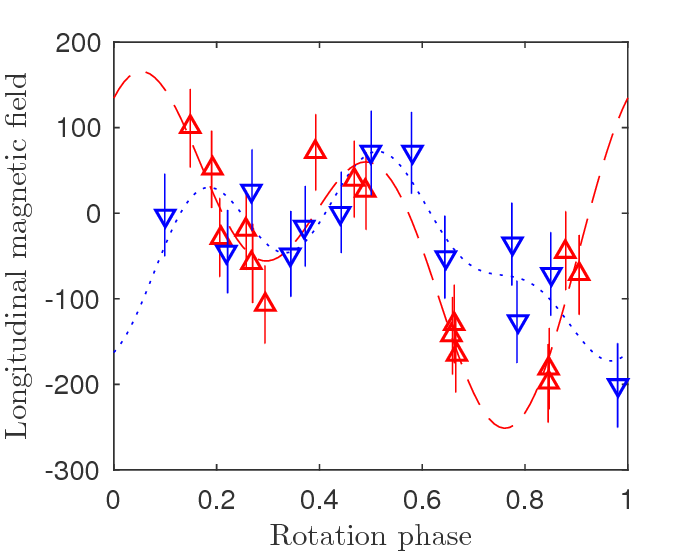}
		\caption{Folded curve of the longitudinal magnetic field against the rotation phase. 2015 Nov (red upward-pointing triangles) data are fitted with the sum of a sine curve and 1 harmonic (red dashed line) and 2016 Jan (blue downward-pointing triangles) data are fitted with the sum of a sine curve and 2 harmonics (blue dotted line).}
		\label{fig:bls}
	\end{figure}
	\begin{figure}
		\centering
		\includegraphics[scale=0.9]{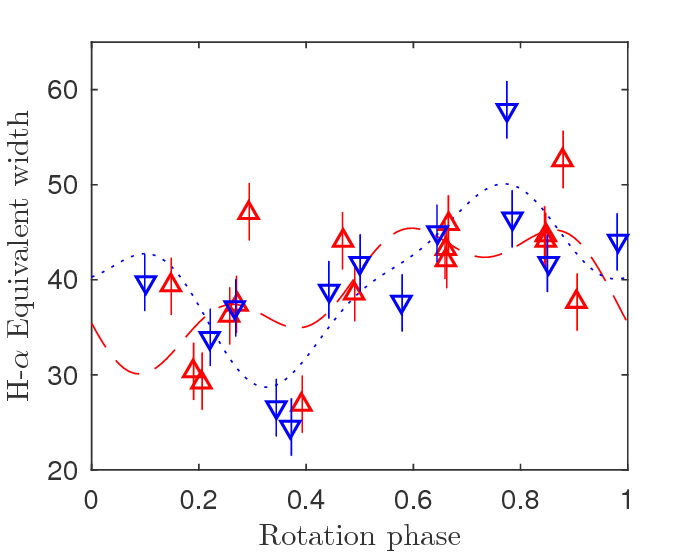}
		\caption{Folded curve of the equivalent width of \hal\ against the rotation phase. 2015 Nov (red upward-pointing triangles) and 2016 Jan (in blue) data are fitted with the sum of a sine curve and 2 harmonics (red dashed line and blue dotted line respectively).}
		\label{fig:hal}
	\end{figure}
	In order to investigate whether the detected periodic RV signal may relate to activity, we plotted periodograms of the longitudinal magnetic field \Bl\ and of the \hal\ emission equivalent width (Fig.~\ref{fig:pbl} and \ref{fig:pha} respectively). Peak frequencies for these proxies are located at periods of 0.7145$\pm$0.0002~d and 0.7132$\pm$0.0001~d respectively, as well as their aliases. Given the surface differential rotation parameters measured for TAP~26 (see Sec.~\ref{sec:par}), the values of their respective periods indicate that the longitudinal field traces an average latitude of 46\degr\ whereas the bulk of \hal\ emission comes from a lower average latitude of 27\degr\ (see Eq.~\ref{eqn:dr}). As opposed to the raw RVs, the rotation period \Prot\ has a higher power than its first harmonic \Prot/2 (Fig.~\ref{fig:per}). No signal is detected at the planet periods found in Sec.~\ref{sec:mpl}.
	
	Plotting phase-folded curves of the longitudinal magnetic field and the \hal\ emission equivalent width (where the x-axis indicates the rotation phase as defined in Eq.~\ref{eq:eph}), in Fig.~\ref{fig:bls} and \ref{fig:hal}, we observe a decrease in the longitudinal magnetic field around phase 0.77 in 2015 Nov and phase 0.97 in 2016 Jan, which correspond approximately to the phases where the dipole pole points towards the Earth (0.73$\pm$0.03 and 0.85$\pm$0.03 respectively), causing \Bl\ to have strong negative values and showing the importance of the dipole in the value of \Bl. Similarly, the increase in emission equivalent width of the \hal\ line between phases 0.6 and 0.9 illustrates the correlation between the lower harmonics of the magnetic field of TAP~26 and this activity proxy. \\
	\vspace{7mm}\\
	% (see also Fig.~\ref{fig:bls} and \ref{fig:hal})
	{\small \it $^1$ Universit\'e de Toulouse, UPS-OMP, IRAP, 14 avenue E.~Belin, Toulouse, F--31400 France\\
		$^2$ CNRS, IRAP~/~UMR 5277, Toulouse, 14 avenue E.~Belin, F--31400 France\\
		$^3$ Department of Physics and Astronomy, York University, Toronto, Ontario L3T 3R1, Canada\\
		$^4$ CFHT Corporation, 65-1238 Mamalahoa Hwy, Kamuela, Hawaii 96743, USA\\
		$^5$ D\'epartement de physique, Universit\'e de Montr\'eal, C.~P.~6128, Succursale Centre-Ville, Montr\'eal, QC, Canada H3C 3J7\\
		$^6$ Crimean Astrophysical Observatory, Nauchny, Crimea 298409\\
		$^7$ ESO, Karl-Schwarzschild-Str 2, D-85748 Garching, Germany\\
		$^8$ SUPA, School of Physics \& Astronomy, Univ. of St Andrews, St Andrews, Scotland KY16 9SS, UK\\
		$^9$ School of Physics, Trinity College Dublin, the University of Dublin, Ireland\\
		$^{10}$ Departamento de Fisica -- ICEx -- UFMG, Av.~Ant\^onio Carlos, 6627, 30270-901 Belo Horizonte, MG, Brazil\\
		$^{11}$ Universit\'e Grenoble Alpes, IPAG, BP 53, F--38041 Grenoble C\'edex 09, France\\
		$^{12}$ CNRS, IPAG~/~UMR 5274, BP 53, F--38041 Grenoble C\'edex 09, France\\
		$^{13}$ Institute of Astronomy and Astrophysics, Academia Sinica, PO Box 23-141, 106, Taipei, Taiwan\\
		$^{14}$ Kavli Institute for Astronomy and Astrophysics, Peking University, Yi He Yuan Lu 5, Haidian Qu, Beijing 100871, China\\
		$^{15}$ LUPM, Universit\'e de Montpellier, CNRS, place E.~Bataillon, F--34095, Montpellier, France}
	
	\FloatBarrier
	\newpage
	\section{Simulations}
	\label{sec:sim}
\begin{figure}
		\centering
		\includegraphics[scale=0.4, angle=-90]{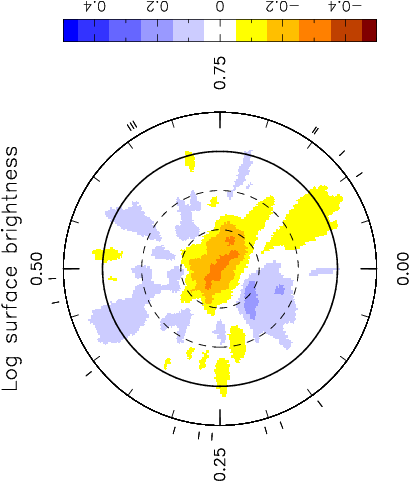}
		\includegraphics[scale=0.4, angle=-90]{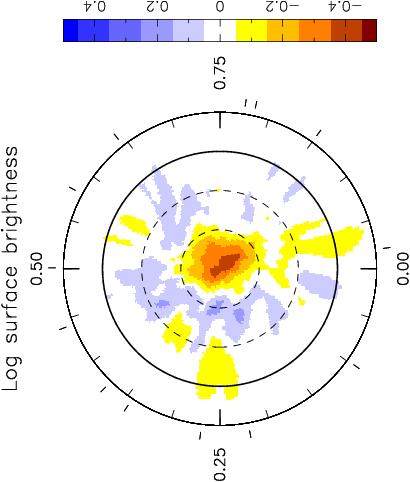}
		\caption[]{Brightness maps reconstructed from the simulation~\#1 data, for the 2015 Nov data subset (top) and the 2016 Jan data subset (bottom). Both maps feature a spot coverage of $\simeq$8\%.}
		\label{fig:sim}
	\end{figure}

	In order to check that the detected planetary signal does not come from the filtering process, we conducted simulations to test our three methods on two different data sets: one where the presence of a planet was input in the simulation (scenario~\#1), and one without any planet (scenario~\#2). Stokes $I$ and $V$ LSD profiles were generated from the brightness and magnetic maps found in Sec.~\ref{sec:mod}, at the same dates of observation as the real data, with a comparable noise level. The added planet signature had the properties of the best fit found with the second method (Sec.~\ref{sec:pla}): $K$=0.154~\kms, \popr=15.29, $\phi$=0.671. Applying ZDI to these data sets, we reconstructed brightness and magnetic maps as in Sec.~\ref{sec:mod}. For both simulations, the maps we found look similar to the ones reconstructed from the real data, with an information loss amounting to 4\% for the spottedness and $\simeq$80~G for the rms magnetic flux, but the main features, such as the polar spot, are recovered. Fig.~\ref{fig:sim} shows the brightness maps for simulation~\#1, at both epochs.
	
\begin{figure*}
		\centering
		\includegraphics[scale=0.48]{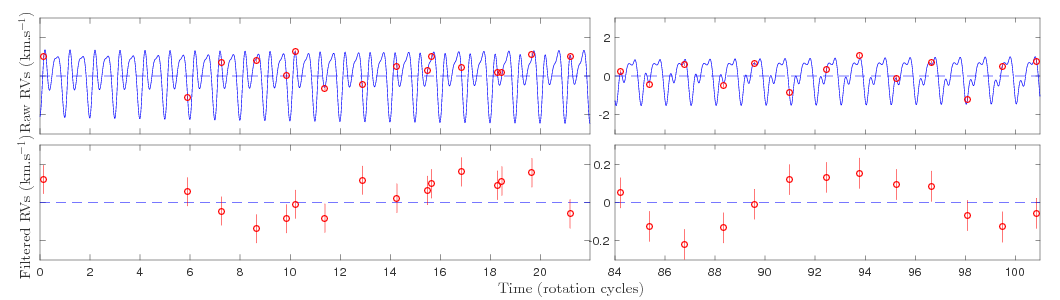}
		\includegraphics[scale=0.55]{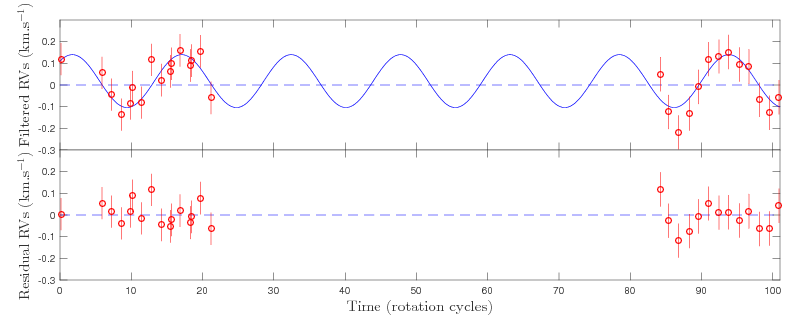}
		\caption[]{Simulation~\#1: raw, filtered and residual RV curves as derived with the method described in \ref{sec:fil}. The residual RVs feature a rms value of 59~\mps.}
		\label{fig:sr1}
	\end{figure*}

\begin{figure*}
		\centering
		\includegraphics[scale=0.48]{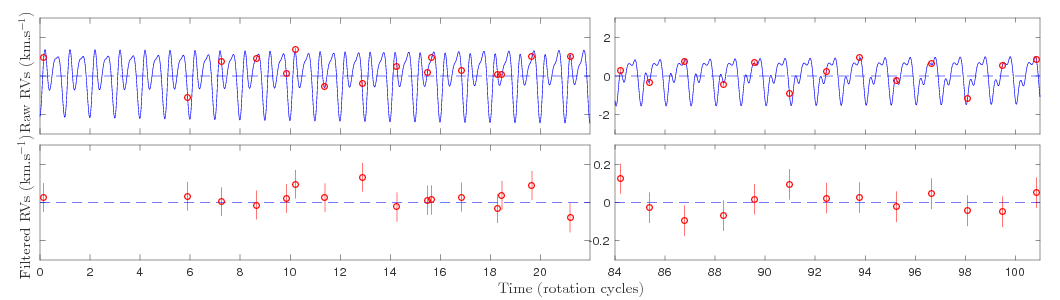}
		\includegraphics[scale=0.55]{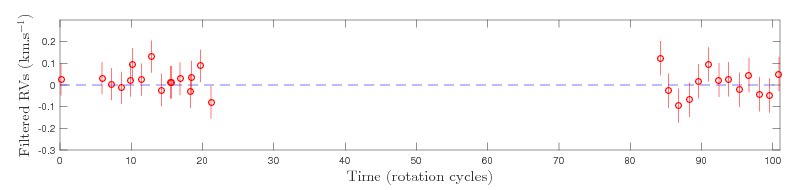}
		\caption[]{Simulation~\#2: raw and filtered RV curves as derived with the method described in \ref{sec:fil}. The filtered RVs feature a rms value of 58~\mps.}
		\label{fig:sr2}
	\end{figure*}

\begin{figure*}
		\centering
		\includegraphics[scale=0.48,angle=-90]{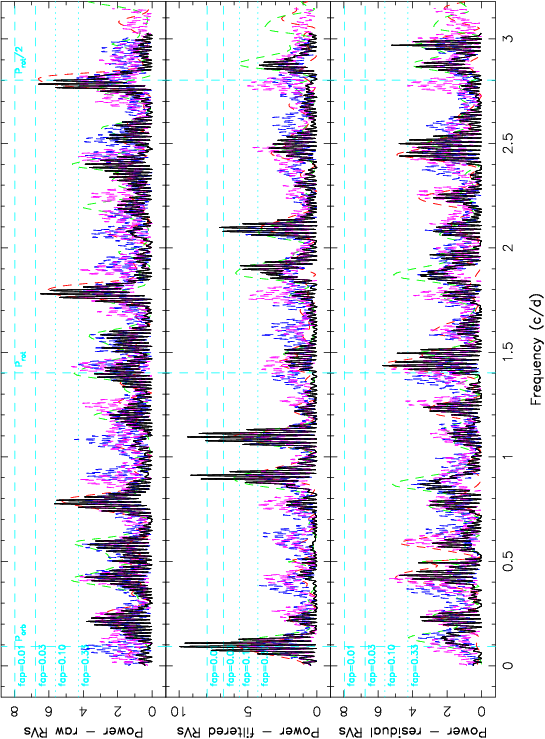}
		\caption[]{Simulation~\#1: periodograms of the raw (top), filtered (middle) and residual (bottom) RVs.}
		\label{fig:pr2a}
	\end{figure*}

\begin{figure*}
		\centering
		\includegraphics[scale=0.48,angle=-90]{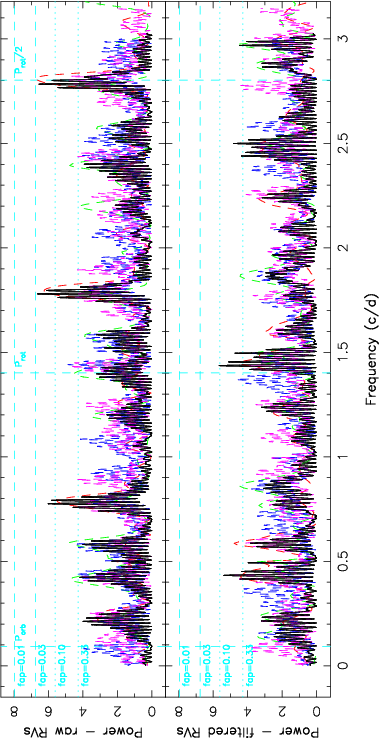}
		\caption[]{Simulation~\#2: periodograms of the raw (top) and filtered (bottom) RVs.}
		\label{fig:pr2b}
	\end{figure*}

	As in Sec.~\ref{sec:fil}, synthetic RV curves are shown in Fig.~\ref{fig:sr1}, \ref{fig:sr2} for simulations~\#1 and \#2 respectively. While a signal is detected in the filtered RVs of simulation~\#1 (rms 107~\mps), no significant signal is detected in the filtered RVs in simulation~\#2 (rms 58~\mps). Table \ref{tab:sim1} summarises the characteristics of the best fit to the filtered RVs for both scenarii, in comparison with a ($K$=0~\kms, RV$_0$=0~\kms) curve. The periodograms of the filtered RV curves, displayed in Fig.~\ref{fig:pr2a} and Fig.~\ref{fig:pr2b}, further confirm this, with simulation~\#2 yielding no significant signal at the frequencies found with our three methods in Sec.~\ref{sec:mpl}.
	
	We note that changing the noise pattern can make the FAP of the filtered RVs highest peak vary between 4\% and 26\% for simulation~\#2 (no planet). It can also change the relative power of the different orbital periods in the filtered RVs for simulation~\#1: most of the time the $\simeq$10.8~d period is recovered as the highest peak with a FAP<0.5\%, but the $\simeq$9.0~d period reaches a smaller \chisqr\ in one case out of five. This sheds light on why the different methods do not always favour the same orbital periods in our analysis.
	
\begin{figure*}
		\centering
		\includegraphics[scale=0.7]{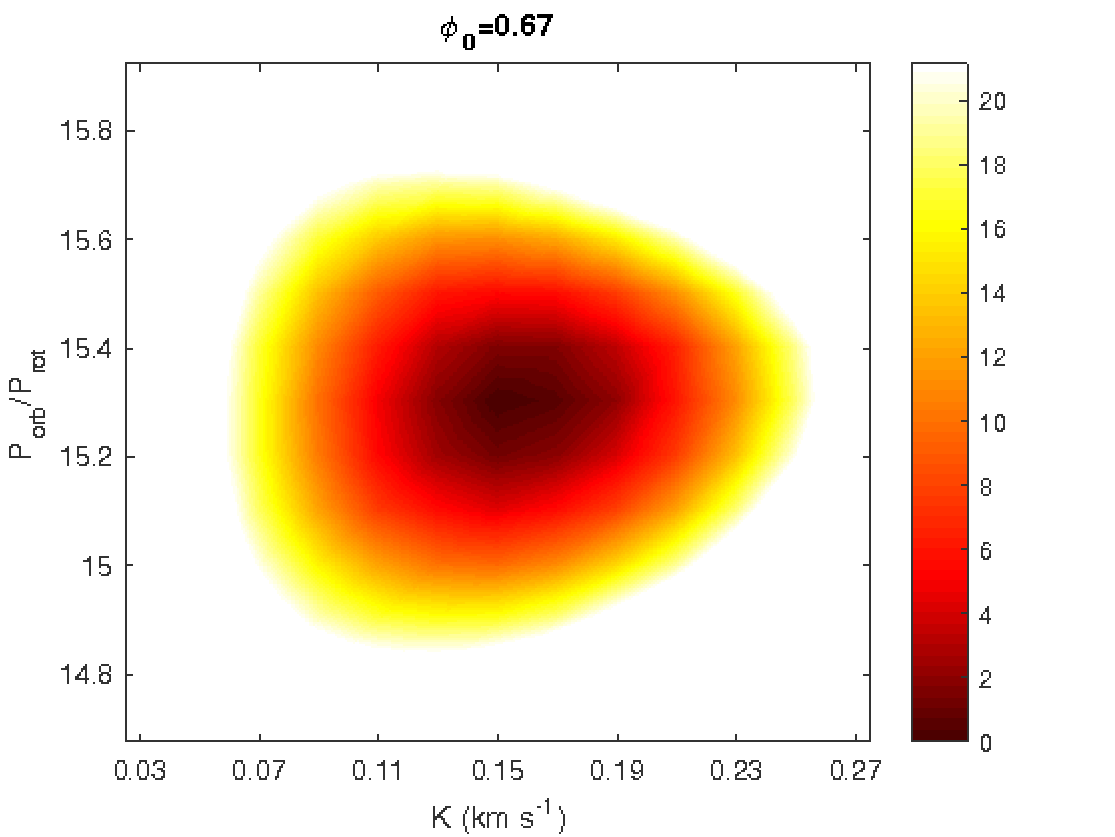}
		\caption{Simulation~\#1: \dchis\ map at $\phi$=0.67 as derived with the method described in \ref{sec:pla}, centered around the minimum \Porb=15.3~\Prot. Parameters values are found to be $K$=0.155$\pm$0.022~\kms, \Porb=15.32$\pm$0.14~\Prot. The minimum value of \chisqr\ is 0.95226}
		\label{fig:spl}
	\end{figure*}

	The second method also recovers the different orbital periods from simulation~\#1, the $\simeq$10.8~d one being the most likely, with a $\Delta\chi^2$ of 7.37 compared to the $\simeq$13.4~d period and 7.67 compared to the $\simeq$9.0~d period. Fig.~\ref{fig:spl} shows the \chisqr\ map around the minimum \Porb$\simeq$10.8~d for $\phi$=0.67 (value at the 3-D local minimum), with the white colour bounding the 99.99\% confidence region. We chose not to apply this method on simulation~\#2 (without planet) because it runs computations for orbital parameters close to the local minima found with the first method, and no such significant minima were found for simulation~\#2.
	\begin{table}
		\centering
		\begin{tabular}{cc}
			\hline
			Scenario~\#1 & Scenario~\#2 \\
			\hline
			$K$=0.122$\pm$0.020~\kms & $K$=0.036$\pm$0.021~\kms \\
			\popr=15.35$\pm$0.16 & \\% \popr=10.05$\pm$0.23 & \\
			$\phi$=0.647$\pm$0.026 & \\% $\phi$=0.367$\pm$0.092 & \\
			RV$_0$=0.018$\pm$0.014~\kms & \\% RV$_0$=0.020$\pm$0.014 & \\
			\chisqr=0.540 & \chisqr=0.436 \\
			\hline
			$K$=0~\kms & $K$=0~\kms \\
			\chisqr=1.893 & \chisqr=0.529 \\
			\hline
			\dchis=39.2 & \dchis=2.7 \\
			$\Delta($\logLr$)$=-16.57 & $\Delta($\logLr$)$=-0.49 \\
			\hline
		\end{tabular}
		\caption{Results found with the 1st method on both simulation datasets. The first column shows the results on the scenario with a planet and the second column shows the results on the scenario without planet. For each, a comparison is made between the best sine fit to the filtered RVs and a fit by a constant value, with the reflex motion semi-amplitude $K$, the orbital period \Porb\ in units of \Prot, the phase of inferior conjunction $\phi$ relative to cycle 11.0, the mean RV RV$_0$ and \chisqr. Differences in $\chi^2$ (summed on 29 data points) and in logarithmic ($\log_e$) likelihood are given in the last row.}
		\label{tab:sim1}
	\end{table}
	\begin{table}
		\centering
		\begin{tabular}{c}
			\hline
			Scenario~\#1 \\
			\hline
			$K$=0.155$\pm$0.022~\kms \\
			\popr=15.32$\pm$0.14 \\
			$\phi$=0.662$\pm$0.036 \\
			\chisqr=0.95226 \\
			\hline
			$K$=0~\kms \\
			\chisqr=0.97529 \\
			\hline
			\dchis=30.0 \\
			$\Delta($\logLr$)$=-10 \\
			\hline
		\end{tabular}
		\caption{Characteristics of the best fit found with the 2nd method on simulation~\#1 (top row), compared to a fit with a no-planet model (i.e. $K$=0~\kms, middle row). Differences in $\chi^2$ (summed on 2581 data points) and in logarithmic ($\log_e$) likelihood are given in the last row.}
		\label{tab:sim2}
	\end{table}
	\begin{table}
		\centering
		\begin{tabular}{cc}
			\hline
			Scenario~\#1 & Scenario~\#2 \\
			\hline
			$K$=0.138$\pm$0.027~\kms & $K$=0.060$\pm$0.052~\kms \\
			\popr=15.31$\pm$0.21 & \\
			$\phi$=0.646$\pm$0.038 & \\
			$\theta_1$=1.14$\pm$0.21~\kms & \\
			$\theta_2$=1.0002$\pm$0.0002~\Prot & \\
			\logLr=-6.20 & \logLr=-5.99 \\
			\hline
			$K$=0~\kms & $K$=0~\kms \\
			\logLr=-21.42 & \logLr=-5.48 \\
			\hline
			$\Delta($\logLr$)$=-15.22 & $\Delta($\logLr$)$=-0.51 \\
			\hline
		\end{tabular}
		\caption{Characteristics of the best fit (first row) found with the 3rd method on simulation~\#1 (left) and \#2 (right), compared to a fit with a no-planet model (i.e. $K$=0~\kms, middle row). Differences in logarithmic ($\log_e$) likelihood are given in the last row.}
		\label{tab:sim3}
	\end{table}
	\begin{figure*}
		\centering
		\includegraphics[scale=0.55, angle=-90]{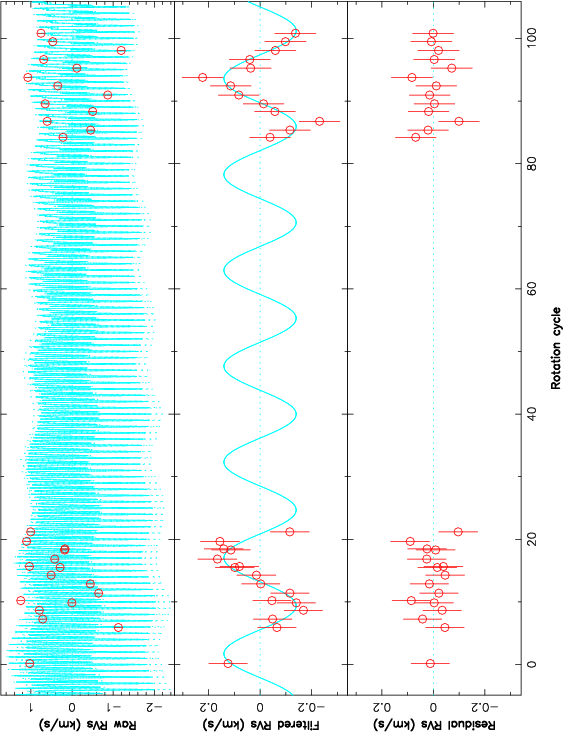}
		\caption{Example of GPR for simulation~\#1. Top: raw RVs (red dots) and GP fit+planet (cyan), middle: curves as derived with the method described in \ref{sec:gau}, for the local minimum \Porb=15.31$\pm$0.21~\Prot, $K$=0.138$\pm$0.027, $\phi$=0.646$\pm$0.038,. The rms of the residual RVs amounts to 47~\mps.}
		\label{fig:sg1}
	\end{figure*}
	\begin{figure*}
		\centering
		\includegraphics[scale=0.6,angle=-90]{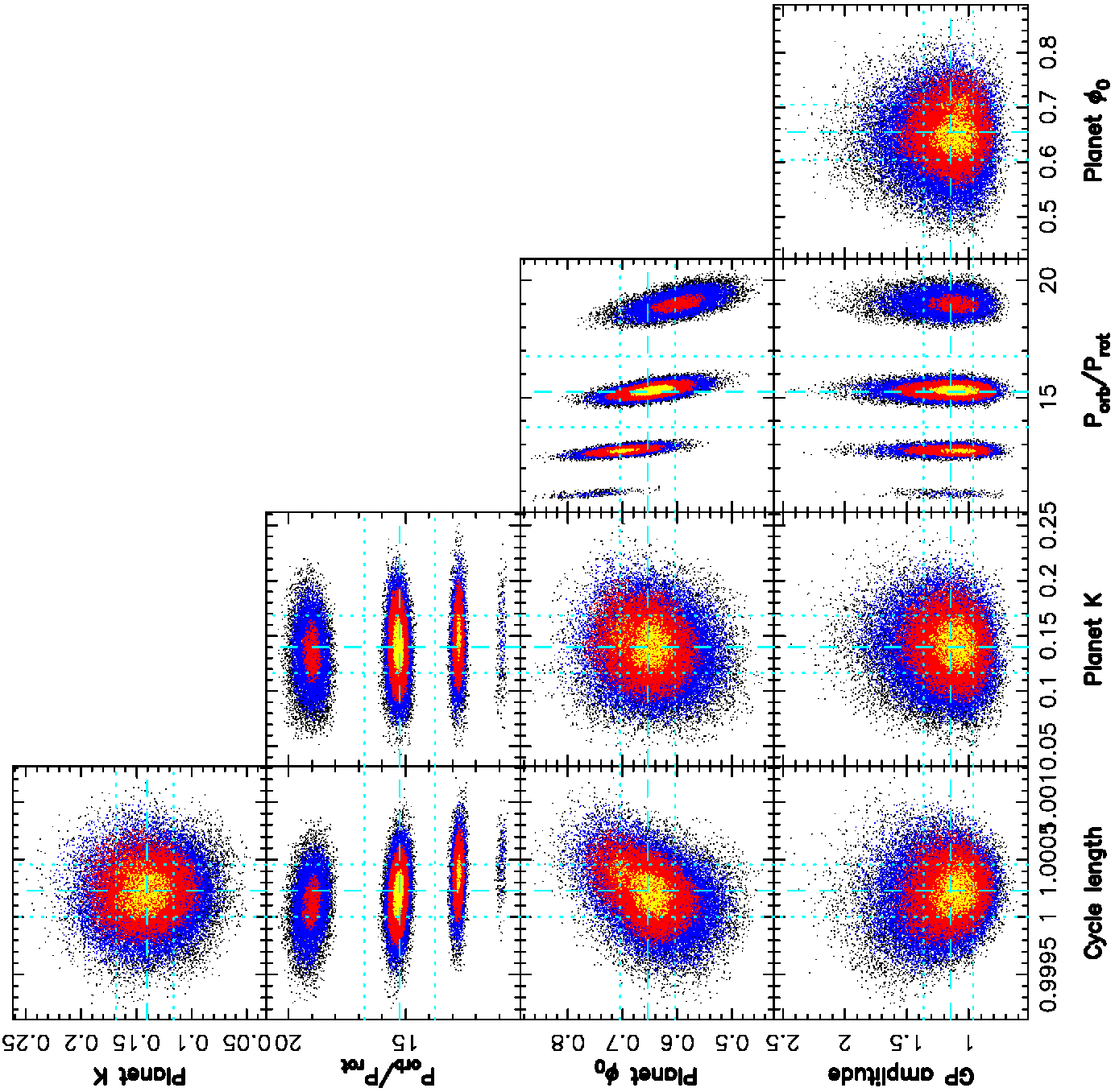}
		\caption{Simulation~\#1: phase plots of the MCMC run as described in \ref{sec:gau}. We find $\theta_1$=1.15$\pm$0.19~\kms, $\theta_2$=1.0002$\pm$0.0002~\Prot, $K$=0.140$\pm$0.026~\kms, and the dominant orbital periods \Porb=15.31$\pm$0.21~\Prot\ and \Porb=12.74$\pm$0.13~\Prot, with the corresponding phases being $\phi$=0.646$\pm$0.038 and $\phi$=0.699$\pm$0.036 respectively.}
		\label{fig:sg2}
	\end{figure*}
	\begin{figure*}
		\centering
		\includegraphics[scale=0.6,angle=-90]{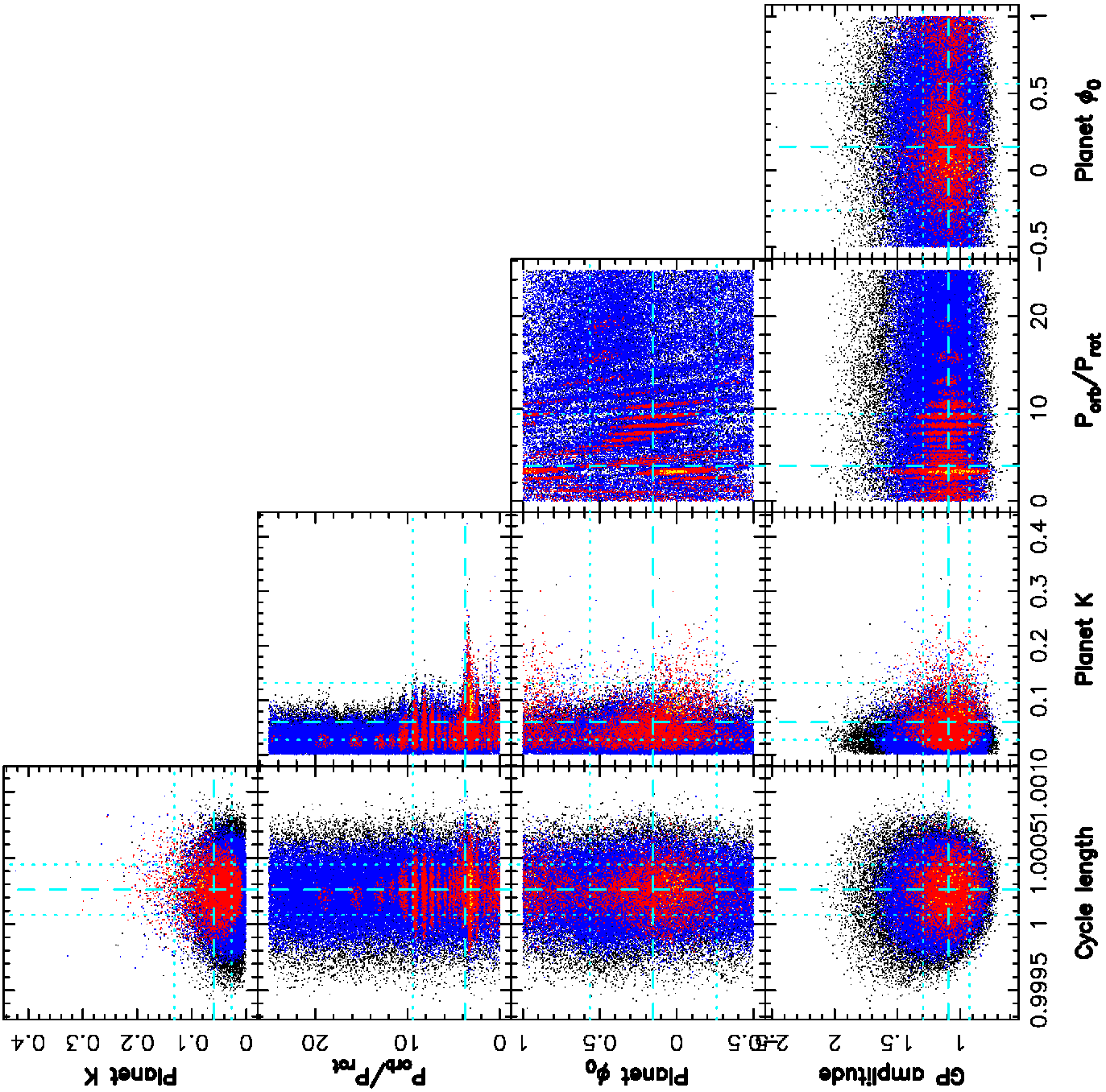}
		\caption{Simulation~\#2: phase plots of the MCMC run as described in \ref{sec:gau}. We find $\theta_1$=1.09$\pm$0.19~\kms, $\theta_2$=1.0003$\pm$0.0002~\Prot\ and $K$=0.060$\pm$0.053~\kms.}
		\label{fig:sg3}
	\end{figure*}
	
	GPR also successfully recovers the input planet period in simulation~\#1, as shown in Fig.~\ref{fig:sg1} on which the raw RVs, the planet signal and the residual RVs at \Porb=15.31\Prot\ are seen. The results of the MCMC runs are shown in the phase plots of Fig.~\ref{fig:sg2} and \ref{fig:sg3}, demonstrating that no orbital period stands out in simulation~\#2, i.e., in the activity jitter synthesised from the reconstructed brightness maps, whereas several orbital periods are detected in simulation~\#1, $\simeq$10.8~d and $\simeq$9.0~d being respectively the most likely and the second most likely, with a Bayes factor of only 1.25 between them. The comparison with a MCMC run where no planet is subtracted is shown in table~\ref{tab:sim3} for both scenarii, demonstrating that, for scenario~\#1, taking a planet into account in the model results in a significant increase in the likelihood of the best fit, whereas it is not the case for scenario~\#2.
	
	We conclude that all three methods enable us to recover the planet signal (scenario~\#1), and that the detected periods in the observational filtered RVs are not artifacts of the numerical process (scenario~\#2). Furthermore, for our particular observation window, the noise pattern can change the relative likelihood of the different detected peaks, as can the choice of the method to use.

	% Don't change these lines
	\bsp	% typesetting comment
	\label{lastpage}
\end{document}